\documentclass[prd,nofootinbib,reprint,twocolumn,showpacs,superscriptaddress,floatfix]{revtex4-1}
\pdfoutput=1 
\usepackage{bm, color} 
\usepackage{amssymb,amsfonts,slashed,amsthm,amsmath, graphicx, soul, empheq, cancel}
\usepackage{hyperref}
\usepackage[caption=false]{subfig}

\begin{document}

\newcommand{\vev}[1]{ \left\langle {#1} \right\rangle }
\newcommand{\bra}[1]{ \langle {#1} | }
\newcommand{\ket}[1]{ | {#1} \rangle }
\newcommand{\eV}{ \ {\rm eV} }
\newcommand{\KeV}{ \ {\rm keV} }
\newcommand{\MeV}{\  {\rm MeV} }
\newcommand{\GeV}{\  {\rm GeV} }
\newcommand{\TeV}{\  {\rm TeV} }
\newcommand{\1}{\mbox{1}\hspace{-0.25em}\mbox{l}}
\newcommand{\Red}[1]{{\color{red} {#1}}}

\newcommand{\lmk}{\left(}  
\newcommand{\rmk}{\right)}
\newcommand{\lkk}{\left[}  
\newcommand{\rkk}{\right]}
\newcommand{\lhk}{\left \{ }  
\newcommand{\rhk}{\right \} }
\newcommand{\del}{\partial}  
\newcommand{\la}{\left\langle} 
\newcommand{\ra}{\right\rangle}
\newcommand{\half}{\frac{1}{2}}

\newcommand{\bea}{\begin{array}}
\newcommand{\eea}{\end{array}}
\newcommand{\beq}{\begin{eqnarray}}
\newcommand{\eeq}{\end{eqnarray}}
\newcommand{\eq}[1]{Eq.~(\ref{#1})}

\newcommand{\dd}{\mathrm{d}}
\newcommand{\Mpl}{M_{\rm Pl}}
\newcommand{\mg}{m_{3/2}}
\newcommand{\abs}[1]{\left\vert {#1} \right\vert}
\newcommand{\mphi}{m_{\phi}}
\newcommand{\Hz}{\ {\rm Hz}}
\newcommand{\for}{\quad \text{for }}
\newcommand{\Min}{\text{Min}}
\newcommand{\Max}{\text{Max}}
\newcommand{\Kahler}{K\"{a}hler }
\newcommand{\cphi}{\varphi}
\newcommand{\Tr}{\text{Tr}}
\newcommand{\diag}{{\rm diag}}

\newcommand{\SUf}{SU(3)_{\rm f}}
\newcommand{\Upq}{U(1)_{\rm PQ}}
\newcommand{\Zpq}{Z^{\rm PQ}_3}
\newcommand{\Cpq}{C_{\rm PQ}}
\newcommand{\ubar}{u^c}
\newcommand{\dbar}{d^c}
\newcommand{\ebar}{e^c}
\newcommand{\nubar}{\nu^c}
\newcommand{\Ndw}{N_{\rm DW}}
\newcommand{\Fpq}{F_{\rm PQ}}
\newcommand{\fpq}{v_{\rm PQ}}
\newcommand{\Br}{{\rm Br}}
\newcommand{\Lag}{\mathcal{L}}
\newcommand{\Lqcd}{\Lambda_{\rm QCD}}

\newcommand{\ji}{j_{\rm inf}} 
\newcommand{\jb}{j_{B-L}} 
\newcommand{\M}{M} 
\newcommand{\im}{{\rm Im} }
\newcommand{\re}{{\rm Re} }

\def\lrf#1#2{ \left(\frac{#1}{#2}\right)}
\def\lrfp#1#2#3{ \left(\frac{#1}{#2} \right)^{#3}}
\def\lrp#1#2{\left( #1 \right)^{#2}}
\def\REF#1{Ref.~\cite{#1}}
\def\SEC#1{Sec.~\ref{#1}}
\def\FIG#1{Fig.~\ref{#1}}
\def\EQ#1{Eq.~(\ref{#1})}
\def\EQS#1{Eqs.~(\ref{#1})}
\def\TEV#1{10^{#1}{\rm\,TeV}}
\def\GEV#1{10^{#1}{\rm\,GeV}}
\def\MEV#1{10^{#1}{\rm\,MeV}}
\def\KEV#1{10^{#1}{\rm\,keV}}
\def\blue#1{\textcolor{blue}{#1}}
\def\red#1{\textcolor{blue}{#1}}

\newcommand{\eff}{\Delta N_{\rm eff}}
\newcommand{\neff}{\Delta N_{\rm eff}}
\newcommand{\cc}{\Omega_\Lambda}
\newcommand{\Mpc}{\ {\rm Mpc}}
\newcommand{\Msolar}{M_\odot}

\def\SYH#1{\textcolor{blue}{[{\bf SYH:} #1]}}
\def\sn#1{\textcolor{red}{#1}}
\def\SN#1{\textcolor{red}{[{\bf SN:} #1]}}
\def\MS#1{\textcolor{blue}{[{\bf MS:} #1]}}

\title{
Bias with a Timer: Axion Domain Wall Decay and Dark Matter
}

\author{Sally Yuxuan Hao}
\affiliation{Tsung-Dao Lee Institute, Shanghai Jiao Tong University, \\
No.~1 Lisuo Road, Pudong New Area, Shanghai 201210, China}
\affiliation{School of Physics and Astronomy, Shanghai Jiao Tong University, \\
800 Dongchuan Road, Shanghai 200240, China}

\author{Shota Nakagawa}
\affiliation{Tsung-Dao Lee Institute, Shanghai Jiao Tong University, \\
No.~1 Lisuo Road, Pudong New Area, Shanghai 201210, China}
\affiliation{School of Physics and Astronomy, Shanghai Jiao Tong University, \\
800 Dongchuan Road, Shanghai 200240, China}

\author{Yuichiro Nakai}
\affiliation{Tsung-Dao Lee Institute, Shanghai Jiao Tong University, \\
No.~1 Lisuo Road, Pudong New Area, Shanghai 201210, China}
\affiliation{School of Physics and Astronomy, Shanghai Jiao Tong University, \\
800 Dongchuan Road, Shanghai 200240, China}

\author{Motoo Suzuki}
\affiliation{SISSA International School for Advanced Studies, Via Bonomea 265, 34136, Trieste, Italy}
\affiliation{INFN, Sezione di Trieste, Via Valerio 2, 34127, Italy}
\affiliation{IFPU, Institute for Fundamental Physics of the Universe, Via Beirut 2, 34014 Trieste, Italy}

\begin{abstract}

We explore the interplay of the post-inflationary QCD axion and a light scalar field
for the axion domain wall decay and dark matter (DM).
The scalar field possesses a nonzero vacuum expectation value (VEV) during inflation,
so that its interaction with the axion effectively serves as an explicit Peccei-Quinn (PQ) violating term. 
At a temperature below the PQ phase transition, the effective PQ violating interaction generates the axion potential
which generally contains multiple degenerate vacua leading to the formation of the axion string-domain wall networks.
The following QCD phase transition provides another contribution to the axion potential making domain walls decay
before they dominate the Universe.
Later, the scalar field starts to relax to the minimum of its potential with a vanishing VEV,
turning off the effective PQ violating interaction so that the axion potential is aligned with the QCD vacuum.
We keep track of the evolution of the axion-scalar system
and discuss the production of the axion DM through the domain wall decay and the (trapped) misalignment.
We find that the string-wall network in some cases can decay due to its structural instability, rather than the volume pressure, 
and the correct axion DM abundance is realized with the decay constant larger than that of the conventional post-inflationary QCD axion without fine tuning.

\end{abstract}

\maketitle

\section{Introduction
\label{sec:introduction}}

The strong CP problem remains one of the most important puzzles in the Standard Model.
Although Quantum Chromodynamics (QCD) allows a CP-violating $\theta$-term in its Lagrangian,
experimental limits on the electric dipole moment (EDM) of the neutron require the physical strong CP phase
$\bar{\theta}$ to be unnaturally small,
$|\bar{\theta}| \lesssim 10^{-10}$
\cite{Baker:2006ts,Pendlebury:2015lrz}.
This tension between the theoretical expectation and observational constraints constitutes the strong CP problem.

A particularly elegant solution to the strong CP problem is provided by the Peccei-Quinn (PQ) mechanism
\cite{Peccei:1977hh}, which promotes the parameter $\bar{\theta}$ to a dynamical field
by introducing a new global $U(1)$ symmetry, spontaneously broken at some high energy scale.
The associated pseudo-Nambu-Goldstone boson, the axion
\cite{Weinberg:1977ma,Wilczek:1977pj}, dynamically relaxes $\bar{\theta}$ to zero.
Remarkably, the axion also serves as a promising candidate for cold dark matter (DM).
Axion DM can be produced through several mechanisms,
depending on the cosmological history and the scale of the PQ symmetry breaking.
The most well-known is the misalignment mechanism \cite{Preskill:1982cy,Abbott:1982af,Dine:1982ah}, in which the axion field,
initially displaced from the minimum of its potential, begins to oscillate when the Hubble parameter drops below the (temperature dependent) axion mass.
Such a coherent oscillation behaves like cold DM.
In the post-inflationary scenario where the PQ symmetry is broken after inflation,
spatial variations in the initial axion field lead to the formation of topological defects,
such as global strings and domain walls,
which also contribute to the axion DM abundance through their decays.

In the present paper, we study a cosmological scenario in which the post-inflationary QCD axion interacts with a light scalar field, resulting in a dynamically controlled mechanism for the axion domain wall decay and DM production.
The introduced scalar field acquires a nonzero vacuum expectation value (VEV) during inflation,
which induces an effective PQ symmetry breaking interaction through its coupling to the axion field.
As the Universe cools after the PQ symmetry is spontaneously broken,
the effective interaction generates an axion potential with multiple degenerate minima,
leading to the formation of a network of the axion strings and domain walls.
Subsequently, the QCD phase transition generates an additional contribution to the axion potential,
which breaks the vacuum degeneracy and triggers the decay of the domain walls
before they dominate the energy density of the Universe.
We point out that the string-wall network in some cases can decay due to its structural instability, rather than the volume pressure.
At a later time, the scalar field evolves to the minimum of its potential, where its VEV vanishes.
This turns off the effective PQ breaking interaction and leaves the axion potential aligned with the CP-conserving QCD vacuum.

Our model is similar to that presented in ref.~\cite{Ibe:2019yew}
where the axion potential, generated by the effective PQ breaking interaction of the scalar field, does not
contain multiple degenerate minima unlike our model, so that DM production through the domain wall decay is negligible.
In addition, ref.~\cite{Ibe:2019yew} have simply assumed 
that the evolution of the axion never backreacts
that of the new scalar field.
Instead, the present paper investigates and presents the evolution of the axion-scalar system
and the resulting axion DM abundance,
which receives contributions from both the domain wall decay and the misalignment mechanism.
We also discuss the importance of the trapping effect of the axion field \cite{Kawasaki:2015lpf,Higaki:2016yqk,Kawasaki:2017xwt,Nakagawa:2020zjr,DiLuzio:2021pxd,DiLuzio:2021gos,Jeong:2022kdr,Nakagawa:2022wwm,DiLuzio:2024fyt}
due to the effective PQ breaking interaction for the estimation of the misalignment contribution.
It is found that the correct DM abundance is realized with the axion decay constant larger than that of the conventional post-inflationary QCD axion without fine tuning.

The rest of the paper is organized as follows.
In section~\ref{sec:scenario}, we present our setup and describe the evolution of the newly introduced scalar field
and the PQ breaking field.
Section~\ref{sec:PQV} discusses the axion potential and the dynamics of the axion-scalar system.
Then, in section~\ref{sec:defect}, we explore the evolution of the axion string-domain wall network.
The DM abundance is estimated in section~\ref{sec:misalignment}.
We outline a possible UV completion in section~\ref{sec:model}.
Section~\ref{sec:Discussion} is devoted to conclusions and discussions.

\section{PQ mechanism with a spectator
\label{sec:scenario}}

We consider a PQ breaking scalar field, which contains the axion for the PQ mechanism, interacting with a light complex scalar,
called a spectator field
\cite{Ibe:2019yew}.
After presenting the setup, we review the evolution of the PQ field and the spectator field.

\subsection{Setup}

Let us introduce a spectator complex scalar field $S$, which interacts with the conventional PQ scalar field $P$
with PQ charge $1$. 
Their potential is given by\footnote{
We generalize the mixing term with $\ell=1$ introduced in ref.~\cite{Ibe:2019yew}.}
\beq
V_{\rm PQ}(P,S) &\supset& \lambda_P\lmk|P|^2-\frac{v_{\rm PQ}^2}{2}\rmk^2 + \frac{1}{(n!)^2} \frac{\lambda_S^2}{\Mpl^{2n-4}}|S|^{2n}\nonumber\\
&+& m_S^2|S|^2 + \lmk\frac{\lambda}{m!\ell! \Mpl^{m+\ell-4}} S^m P^\ell +{\rm h.c.}\rmk,\nonumber\\
\label{VPQ}
\eeq
where $v_{\rm PQ}$ represents the breaking scale of the PQ symmetry, $m_S$ is a mass parameter, $\lambda_P,\lambda_S,\lambda$ are dimensionless coupling coefficients, and $\Mpl\equiv1/\sqrt{8\pi G}$ with the Newton constant $G$.
The mass parameter $m_S$ is assumed to be small so that it becomes relevant after the QCD phase transition.
One can assign the PQ charge $-\ell/m~(m,\ell\in \mathbb{N})$ for the spectator field $S$ to respect the PQ symmetry.
Throughout the paper, we assume that $n~(\in\mathbb{N})$ is larger than or equal to $5$.
We will discuss how other lower dimensional operators are suppressed in \SEC{sec:model}.
When $S$ has a large field value, the evolution of $P$ can be affected significantly due to the mixing interaction.

The PQ breaking field $P$ and the spectator field $S$ interact with the inflaton field $\phi$,
\beq
V_{\rm PQ} \supset V_{\rm inf}(\phi) + \frac{c_P}{3}\frac{V_{\rm inf}(\phi)}{\Mpl^2}|P|^2 - \frac{c_S}{3}\frac{V_{\rm inf}(\phi)}{\Mpl^2}|S|^2 \ .
\label{Vinf}
\eeq
Here $V_{\rm inf}$ is the inflaton potential, and $c_P, c_S$ are taken to be  positive constants.
The interaction terms with the inflaton respectively lead to positive and negative mass terms to $P$ and $S$ during inflation.

We ignore the finite temperature correction to the potential of $S$. 
This can be justified when the inflaton decays dominantly to the Standard Model sector,
and $S$ is never thermalized by the tiny mixing with $P$.
We also note that this scenario is applicable to any specific axion models,
such as KSVZ~\cite{Kim:1979if,Shifman:1979if} and DFSZ~\cite{Zhitnitsky:1980tq,Dine:1981rt} models.

\subsection{Evolution of the spectator field}

The interaction terms in Eqs.~\eqref{VPQ}, \eqref{Vinf} induce a large expectation value $\langle S\rangle$
from inflationary epoch to radiation-dominated era \cite{Harigaya:2015hha,Ibe:2019yew}.
Following ref.~\cite{Ibe:2019yew}, we summarize the analytic formula of $\langle S\rangle$ for each epoch.

During inflation, $V_{\rm inf}\simeq 3H_{\rm inf}^2\Mpl^2$ with $H_{\rm inf}$ the Hubble parameter during inflation,
and $S$ has a negative Hubble mass term,
\beq
V(S) \simeq \frac{1}{(n!)^2} \frac{\lambda_S^2}{\Mpl^{2n-4}}|S|^{2n} - c_S H_{\rm inf}^2 |S|^2 \ .
\eeq
Here we assume that the mixing term is negligibly small compared to these terms.
The expectation value is estimated at the potential minimum, 
\beq
\langle S_{\rm inf} \rangle \simeq \lmk\sqrt{\frac{c_S}{n}}\frac{n!}{\lambda_S}\rmk^{\frac{1}{n-1}}\lmk\frac{H_{\rm inf}}{\Mpl}\rmk^{\frac{1}{n-1}}\Mpl \ .
\label{Sinf}
\eeq
For example, $\langle S_{\rm inf}\rangle\sim \Mpl$ for $H_{\rm inf}=10^{12}\GeV$, $n=6$, $c_S=1$,
and $\lambda_S=10^{-4}$, which means that $\lambda_S\gtrsim10^{-4}$ for a sub-Planckian VEV of $S$.
The large field value during inflation suppresses the isocurvature bound.
As will be discussed in \SEC{sec:misalignment}, we can also take $\lambda_S\sim 1$ to suppress the isocurvature bound sufficiently, but as a more conservative value, we take $\lambda_S=10^{-4}$ throughout the paper, which does not change any overall consequences.

After inflation, the inflaton oscillation dominates the energy density in the Universe.
Since the inflaton oscillates with a time scale shorter than the Hubble time,
$S$ evolves with the time-averaged inflaton potential, 
\beq
\bar{V}_{\rm inf} \simeq \frac{3}{2}H^2\Mpl^2 \ .
\eeq
The equation of motion for $S$ is given by
\beq
\ddot{|S|} +3H\dot{|S|} +\frac{n\lambda_S^2}{(n!)^2\Mpl^{2n-4}}|S|^{2n-1} -\frac{c_S}{2}H^2|S|=0 \ ,
\eeq
where the dot denotes the derivative with respect to cosmic time $t$. 
We neglect the phase direction of $S$ approximately, as long as $m\gg\ell$, so that any motion is only slightly induced in the phase direction.
Following the method described in ref.~\cite{Ibe:2019yew},
we obtain a large expectation value of $S$ during inflaton domination,
\beq
\langle |S_{\rm ID} |\rangle \simeq \lmk\sqrt{\frac{c_S}{n}}\frac{n!}{\lambda_S}\rmk^{\frac{1}{n-1}}\lmk\frac{H}{\Mpl}\rmk^{\frac{1}{n-1}}\Mpl \ .
\label{SID}
\eeq
This behavior is known as the scaling solution \cite{Liddle:1998xm}.

After reheating, the Universe undergoes the radiation dominated era.
Since $S$ no longer feels the inflaton coupling, the equation of motion of $S$ is given by
\beq
\ddot{|S|} +3H\dot{|S|} +\frac{n\lambda_S^2}{(n!)^2\Mpl^{2n-4}}|S|^{2n-1}=0 \ .
\eeq
We then obtain a value of $S$ during the radiation dominated era for $n\geq6$
\cite{Harigaya:2015hha},
\beq
\langle |S_{\rm RD}| \rangle \simeq \left[\frac{2(n-3)(n!)^2}{n(n-1)^2\lambda_S^2}\right]^{\frac{1}{2(n-1)}}\lmk\frac{H}{\Mpl}\rmk^{\frac{1}{n-1}}\Mpl \ .
\label{SRD}
\eeq
The spectator field gradually rolls down on the potential, which would not be seen in the presence of a quadratic potential.
This is because the driving force competes with the Hubble friction.
Note that the case of $n=5$ also leads to a large value of $S$,
but its behavior is different and more complicated.
Thus, we choose $n=6$ as a reference value throughout the present paper.\footnote{Likewise, $S$ obeys the scaling solution in the matter-dominated Universe, given by $\langle S_{\rm MD}\rangle \simeq \langle S_{\rm RD}\rangle\times(9(n-2)/8(n-3))^{1/(2n-2)}$, which corresponds to the case with $m_S\lesssim10^{-28}\eV$.}

Throughout the history in the Universe until the oscillation of $S$ starts,
the spectator field has a large field value (\ref{Sinf}), (\ref{SID}), and (\ref{SRD}),
leading to a large PQ breaking potential via the mixing term in Eq.~\eqref{VPQ}.
When $m_S\sim H$, $S$ starts to oscillate around its origin, so that
the PQ breaking effect is suppressed and
becomes small enough to evade the experimental bound of the neutron EDM.
Since we focus on the case where the PQ breaking is effective in the axion evolution until the QCD confinement, 
we assume the range of the mass parameter,
\beq
m_S \lesssim \sqrt{\frac{\pi^2g_*}{90}}\frac{\Lambda_{\rm QCD}^2}{\Mpl} 
\simeq 3\times 10^{-11}\eV \ ,
\label{lateSosc}
\eeq
where $\Lambda_{\rm QCD} \simeq 150\MeV$ denotes the dynamical scale of QCD.
Note that the abundance of coherent oscillation of $S$ is negligibly small for this mass range.
In fact, assuming that the onset of the oscillation is right after the QCD scale, the current abundance of $S$ is estimated as
\beq
\rho_S/\rho_{\rm DM} \approx \frac{m_S^2|S(\Lambda_{\rm QCD})|^2}{H_0^2\Mpl^2}\frac{s_0}{s(\Lambda_{\rm QCD})}\sim \mathcal{O}(10^{-4}),
\eeq
where $s$ is the entropy density, and $H_0,s_0$ denote the present value of the Hubble parameter and the entropy density.

\subsection{Evolution of the PQ field}

During inflation, the potential of $P$ is described by
\beq
V_P &=& (c_P H_{\rm inf}^2-\lambda_P v_{\rm PQ}^2)|P|^2 + \lambda_P|P|^4\nonumber\\
&+& \frac{\lambda}{m!\ell!\Mpl^{m+\ell-4}}\langle S_{\rm inf}\rangle^m P^\ell + {\rm c.c.} \ ,
\eeq
where the quadratic term is positive, assuming $H_{\rm inf}\gg v_{\rm PQ}$, and thus,
the minimum of $P$ depends on the latter two terms.\footnote{
There is a single minimum at the origin, $\langle P_{\rm inf}\rangle\simeq0$, for even values of $\ell$.
On the other hand, when $\ell$ is odd, the analysis depends on $\ell$.
For $\ell=1$, a single minimum deviates from the origin due to the linear term.
For $\ell=3$, there are two minima.
For $\ell=5,7,9,\cdots$, the potential has a single minimum but is not bounded from below,
which requires an additional higher dimensional $(>\ell)$ potential.
Here, the phase direction is assumed to be stabilized at the origin for simplicity.}

When the Universe reheats, $P$ acquires a thermal potential due to its coupling to the thermal bath.
For instance, in the KSVZ model, it is given by the coupling to PQ-charged quarks.
As the Universe cools, the effect of the thermal potential gets smaller, and
at $T\sim v_{\rm PQ}$, the PQ symmetry is spontaneously broken, so that an axion appears in the low-energy effective theory.

\section{Effects of PQ breaking
\label{sec:PQV}}

We now discuss the effective axion potential generated from the mixing term in Eq.~\eqref{VPQ}
and the non-perturbative QCD effect,
and investigate the dynamics of the axion-spectator field system.

\subsection{Effective axion potential}

The mixing interaction gives rise to an effective potential of the axion, as long as $S$ keeps a large VEV. 
After the PQ symmetry is spontaneously broken, we parameterize the PQ field and the spectator field by\footnote{Recently, a similar scenario was discussed in \cite{Lee:2025zpn}, where the authors also considered two axion setup but the evolution of the radial mode is irrelevant.}
\beq
P = \frac{v_{\rm PQ}}{\sqrt{2}}e^{ia/v_{\rm PQ}} \ ,~~~
S = \frac{\chi}{\sqrt{2}}e^{ib/\chi} \ , 
\label{parameterize}
\eeq
to obtain the potential, 
\beq
V_{\cancel{\rm PQ}} \simeq -\frac{1}{\ell^2} m_{\cancel{\rm PQ}}^2v_{\rm PQ}^2
\cos\lmk\ell\frac{a}{v_{\rm PQ}} + m \frac{b}{\chi} +\delta\rmk.
\label{PQVpotential}
\eeq
Here $\delta$ denotes a constant phase of $\lambda$, and 
the corresponding mass for the axion is defined as
\beq
m^2_{\cancel{\rm PQ}}(T) \simeq \frac{|\lambda|\ell^2}{2^{\ell/2-1}m!\ell!} \frac{\langle S_{\rm RD}\rangle^m v_{\rm PQ}^{\ell-2}}{\Mpl^{m+\ell-4}} \ .
\label{mPQ}
\eeq
Note that the Lagrangian is PQ symmetric, but the potential effectively breaks the PQ symmetry for a nonzero value of $S$.

\begin{figure*}[t!]
\centering
\includegraphics[width=8cm]{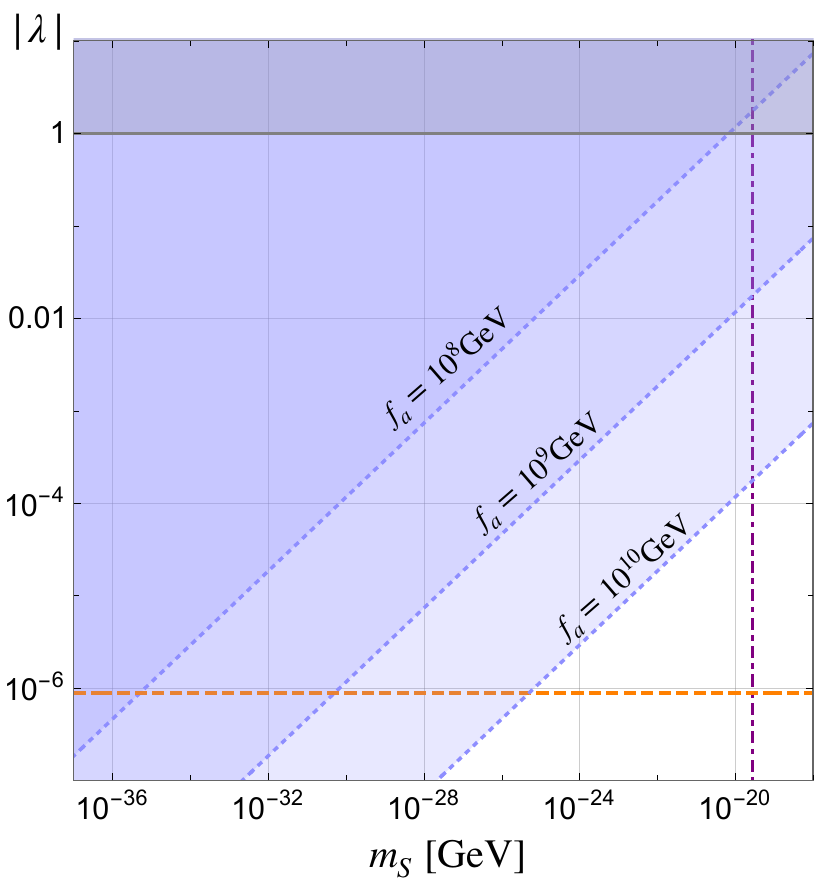}
\caption{
The parameter space in the $(m_S,|\lambda|)$ plane.
The blue shaded regions show the limit (\ref{backreaction}) for various $f_a$ (the denser region, the smaller $f_a$).
The backreaction is not negligible in the region above each line.
The orange dashed lines correspond to the threshold $T_{\rm osc}>T_{\rm osc}^{\rm (conv)}$.
The region above each line satisfies the threshold. 
The purple dot-dashed line represents the upper bound on $m_S$ of our interest (\ref{lateSosc}).
The gray shaded region corresponds to $|\lambda|>1$.
For $\lambda_S=10^{-4}$, we set
$(N_{\rm DW},\ell,m,n)=(3,2,10,6)$.
}
\label{fig:focus}
\end{figure*}

In addition to the mixing interaction,
the axion acquires a potential via non-perturbative effects of QCD, 
\beq
V_{\rm QCD}(a) = m_a^2(T) f_a^2 \lmk1-\cos\frac{a}{f_a}\rmk,
\label{VQCD}
\eeq
where we define $f_a\equiv v_{\rm PQ}/N_{\rm DW}$ as the axion decay constant with $N_{\rm DW}$ the domain wall number.
We refer to the lattice results \cite{Borsanyi:2016ksw} for the temperature-dependent axion mass, $m_{a}(T)$ for $T\gtrsim\Lambda_{\rm QCD}$, given by
\beq
m_{a}(T)\simeq m_{a,0} \lmk\frac{T}{\Lambda_{\rm QCD}}\rmk^{-\tilde{b}} ,
\eeq
with a numerical exponent $\tilde{b}=3.92$.\footnote{We use the same function of the axion mass
as that of ref.~\cite{Jeong:2022kdr}.}
Here $m_{a,0}$ is defined as the zero temperature mass for $T\lesssim \Lambda_{\rm QCD}$.
By the chiral perturbation theory \cite{GrillidiCortona:2015jxo,Gorghetto:2018ocs}, the mass is estimated as
\beq
m_{a,0}\simeq 0.57 \, {\rm meV} \lmk\frac{10^{10}\GeV}{f_a}\rmk.
\eeq

\begin{figure*}[t!]
\begin{minipage}[t]{16.5cm}
\centering
\includegraphics[width=7.9cm]{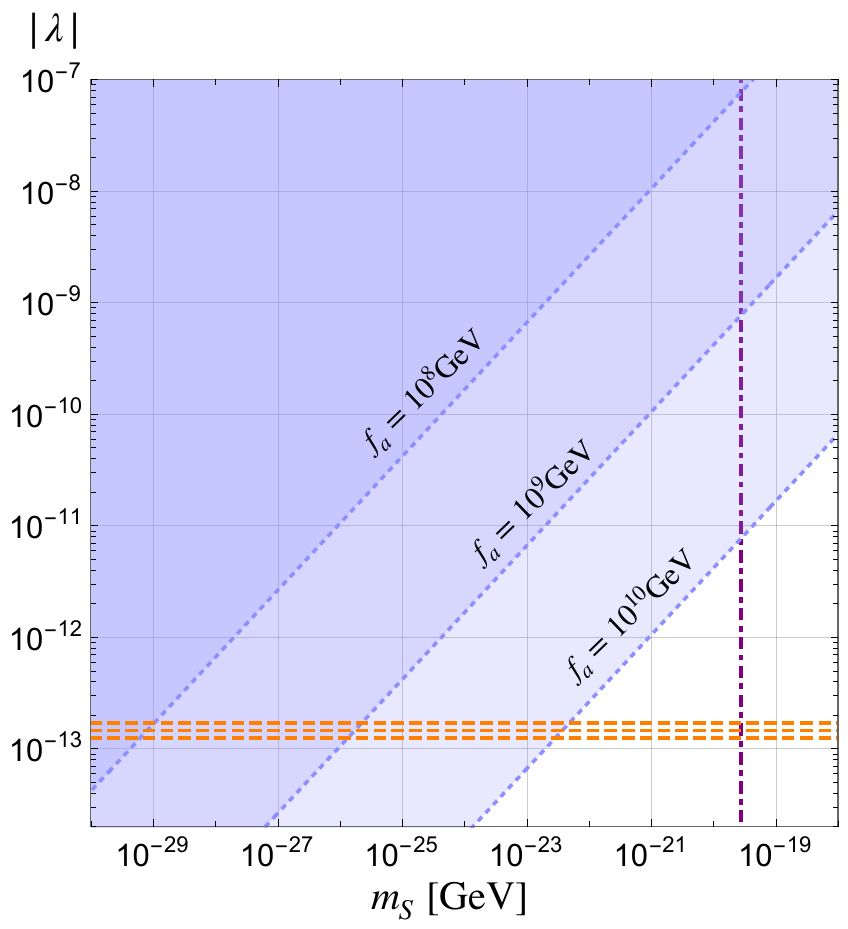}
\hspace{5mm}
\centering
\includegraphics[width=7.9cm]{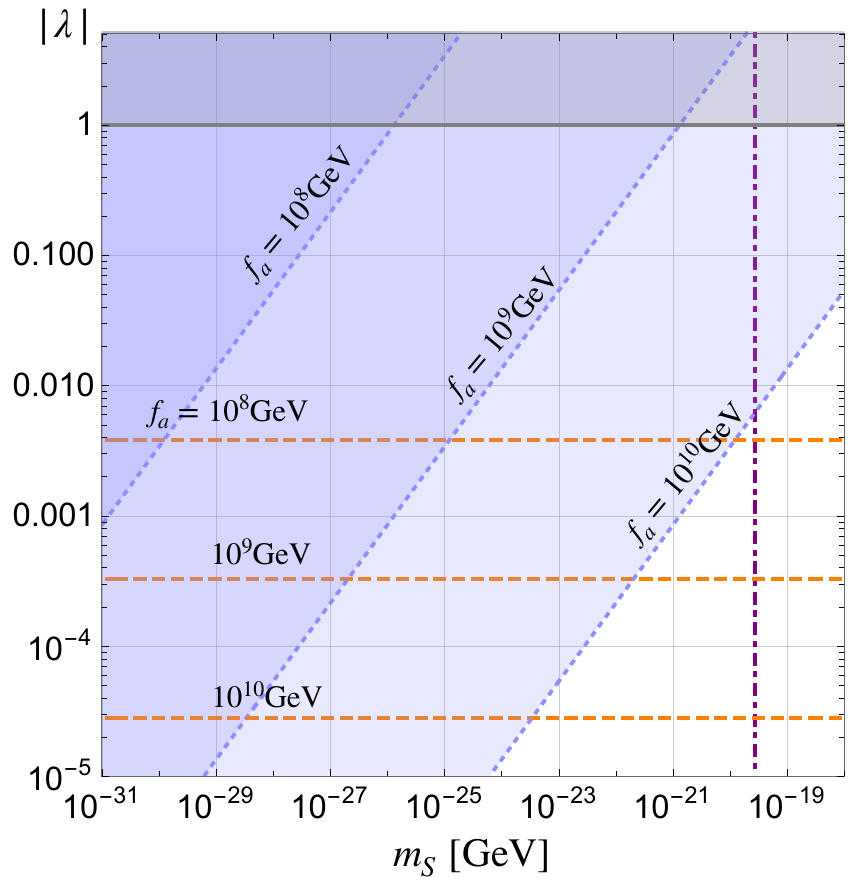}
\end{minipage}
\caption{
The parameter space in the $(m_S,|\lambda|)$ plane.
For $\lambda_S=10^{-4}$, we set
$(N_{\rm DW},\ell,m,n)=(3,2,9,6)$ and $(2,3,9,6)$ in the left and right panels, respectively.
See \FIG{fig:focus} for the description of lines.
Note that the orange dashed lines in the left panel correspond to $f_a=10^8,10^9,10^{10}\GeV$ from the above.
}
\label{fig:focus2}
\end{figure*}

Since the potential (\ref{PQVpotential}) can be dominant over the usual one $(\ref{VQCD})$ before the QCD confinement,
it affects the axion dynamics in a nontrivial way. 
Our focus is on a scenario that the QCD axion starts to oscillate earlier than the conventional onset of oscillation,
$T_{\rm osc}\gg T_{\rm osc}^{\rm (conv)}$
which is estimated by $H\sim m_{a}$,
\beq
T_{\rm osc}^{\rm (conv)} \simeq 2.6\GeV \lmk\frac{g_*}{80}\rmk^{-0.084} \lmk\frac{f_a}{10^{10}\GeV}\rmk^{-0.17}.
\eeq
Due to the dependence of $m_{\cancel{\rm PQ}}$ on $m$, we consider two cases: (i) $m\geq2n-2$ and (ii) $m\leq2n-3$.

\vspace{3mm}

\noindent
(i) $m\geq2n-2$\\[1ex]
Since $m_{\cancel{\rm PQ}}\propto H^{m/(2n-2)}$, it decreases faster than or at the same rate with the Hubble parameter.
The condition $T_{\rm osc}\gg T_{\rm osc}^{\rm (conv)}$ is satisfied, as long as $H\lesssim m_{\cancel{\rm PQ}}$ at the PQ breaking scale so that we have
\beq
|\lambda|>\frac{2^{\ell/2-1}m!\ell!}{\ell^2}\frac{\pi^2g_*}{90}\lmk\frac{v_{\rm PQ}}{\Mpl}\rmk^{6-\ell}\lmk\frac{\langle S_{\rm RD}\rangle}{\Mpl}\rmk^{-m}.
\eeq
Thus, $T_{\rm osc}\simeq v_{\rm PQ}$.
Note that a larger exponent $m$ of the mixing term requires a larger value of $|\lambda|$ due to this constraint.
In fact, we find that there is no allowed region for $m\geq2n$, 
i.e., $|\lambda|$ is much larger than one for $n=6$, when taking $\ell>1$ we are interested in.
In the next section, we will see that the case of $m=2n-1$ is also restricted.
Then, we only consider $m=2n-2$.
As will be discussed later, the domain wall density in the scaling scheme is estimated by $\rho_{\rm wall}\sim m_{\cancel{\rm PQ}}v_{\rm PQ}^2 H\propto H^{m/(2n-2)+1}$, so that the domain wall density dilutes at the same rate as the radiation for $m=2n-2$.

\vspace{3mm}

\noindent
(ii) $m\leq2n-3$\\[1ex]
In this case, $m_{\cancel{\rm PQ}}$ decreases more slowly than $H$.
The oscillation temperature is then given by
\beq
T_{\rm osc} &\simeq& 20\TeV \lmk\frac{106.75}{g_*}\rmk^{\frac{1}{4}} \lmk\frac{f_a}{10^{10}\GeV}\rmk^{\frac{5}{2}}\nonumber\\
&\times& \lmk\frac{N_{\rm DW}}{2}\rmk^{\frac{5}{2}} \lmk\frac{|\lambda|}{10^{-3}}\rmk^{\frac{5}{2}}\lmk\frac{\lambda_S}{10^{-4}}\rmk^{-\frac{9}{2}}  , 
\eeq
for $T_{\rm osc}<v_{\rm PQ}$, otherwise $T_{\rm osc}=v_{\rm PQ}$.
Here we set $\ell=3$, $m=9$, and $n=6$.
Solving $T_{\rm osc}\gg T_{\rm osc}^{\rm (conv)}$, we obtain the lower bound on $|\lambda|$.
Since the domain wall density is proportional to $H^{m/(2n-2)+1}$ with the exponent $m/(2n-2)+1$ smaller than but close to $2$, the domain wall cannot dominate the Universe until the QCD scale.

\subsection{Backreaction}

So far, we have implicitly assumed that the evolution of $P$ never backreacts that of $S$
to simplify the analysis \cite{Ibe:2019yew}, which is justified by imposing a condition,
\beq
\frac{1}{(n!)^2} \frac{\lambda_S^2}{\Mpl^{2n-4}}|S|^{2n} > \frac{|\lambda|}{m!\ell! \Mpl^{m+\ell-4}} |S|^m v_{\rm PQ}^\ell \ ,
\label{rhoconst}
\eeq
at any temperature from the spontaneous breaking of the PQ symmetry to the onset of oscillation of $S$.
We then obtain the upper bound on $|\lambda|$,
\beq
|\lambda| < \frac{m!\ell!\lambda_S^2}{(n!)^2}\lmk\frac{\Mpl}{v_{\rm PQ}}\rmk^\ell \left[\frac{2(n-3)(n!)^2}{n(n-1)^2\lambda_S^2} \frac{m_S^2}{\Mpl^2}\right]^{\frac{2n-m}{2(n-1)}}.
\label{backreaction}
\eeq
We note that this bound is not a constraint but a limit of the analysis.

Let us summarize the parameter space $(m_S,|\lambda|)$ with the constraints of $T_{\rm osc}>T_{\rm osc}^{(\rm conv)}$ and the backreaction (\ref{backreaction}).
For $\lambda_S=10^{-4}$, we set $(N_{\rm DW}, \ell,m,n)=(3,2,10,6)$ in \FIG{fig:focus} and $(N_{\rm DW},\ell,m,n)=(3,2,9,6),~(2,3,9,6)$ in the left and right panels of \FIG{fig:focus2}, respectively.
To develop a global minimum, we choose $N_{\rm DW}$ and $\ell$ to be co-prime throughout the present paper.
As will be discussed in \SEC{sec:defect}, this guarantees that the potential term $V_{\cancel{\rm PQ}}$ explicitly breaks the degeneracy, leading to instability of domain walls.
We note that the (mild) degeneracy of the orange lines for $\ell=2$ is attributed to the independence of $m_{\cancel{\rm PQ}}^2$ on $v_{\rm PQ}$.
In addition, it is possible to take many other sets of $(\ell,m,n)$.
Since larger $\ell$ and $m$ give a more significant Planck suppression, the maximum value of $\ell$ depends on $m$ by assuming not too large $|\lambda|$.
Let us enumerate possible parameter sets as follows: 
\beq
(\ell,m,n)&\ni&(1,10,6),(2,10,6)\nonumber\\
&&(1,9,6),(2,9,6),(3,9,6)\nonumber\\
&&(1,8,6),(2,8,6),(3,8,6),(4,8,6)\nonumber\\
&&(1,7,6),(2,7,6),(3,7,6),(4,7,6),(5,7,6).\nonumber
\eeq
One can see that we need a very small value of $|\lambda|$ unless we consider the maximum value of $\ell$ for some $m$ (e.g. see the left panel of \FIG{fig:focus2}), and such a small value of $|\lambda|$ should be addressed in a UV model, which will be discussed in \SEC{sec:model}.
Furthermore, one can see from \FIG{fig:focus} and \FIG{fig:focus2} that these constraints put an upper bound on $f_a$ (typically we find $f_a\lesssim 10^{10}\GeV$ for $m\leq2n-2$).
The backreaction becomes more effective for larger $f_a$, while the condition
$T_{\rm osc}>T_{\rm osc}^{(\rm conv)}$ mildly depends on $f_a$.

In ref.~\cite{Ibe:2019yew}, the region with non-negligible backreaction was not explicitly presented.
In the next subsection, we will clarify what the backreaction is and its several phenomenological aspects by following the full equations of motion in the system.

\begin{figure*}[t!]
\begin{minipage}[t]{16.5cm}
\centering
\includegraphics[width=7.8cm]{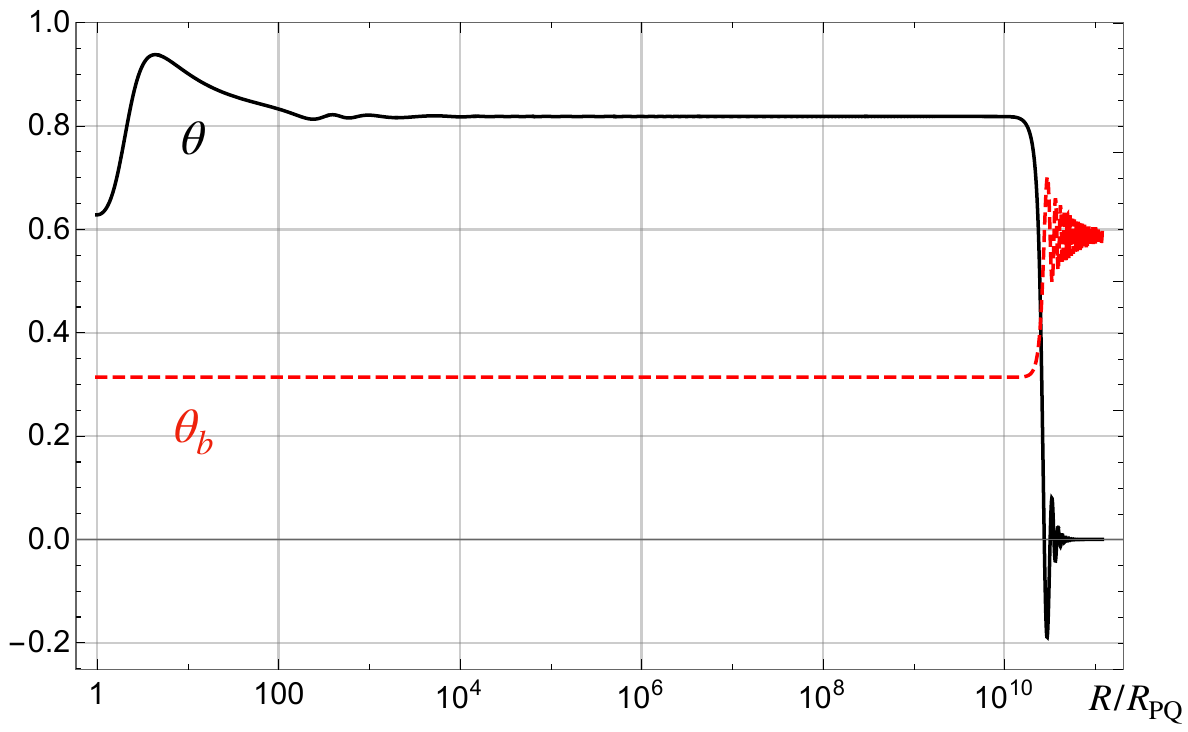}
\hspace{0mm}
\centering
\includegraphics[width=8.4cm]{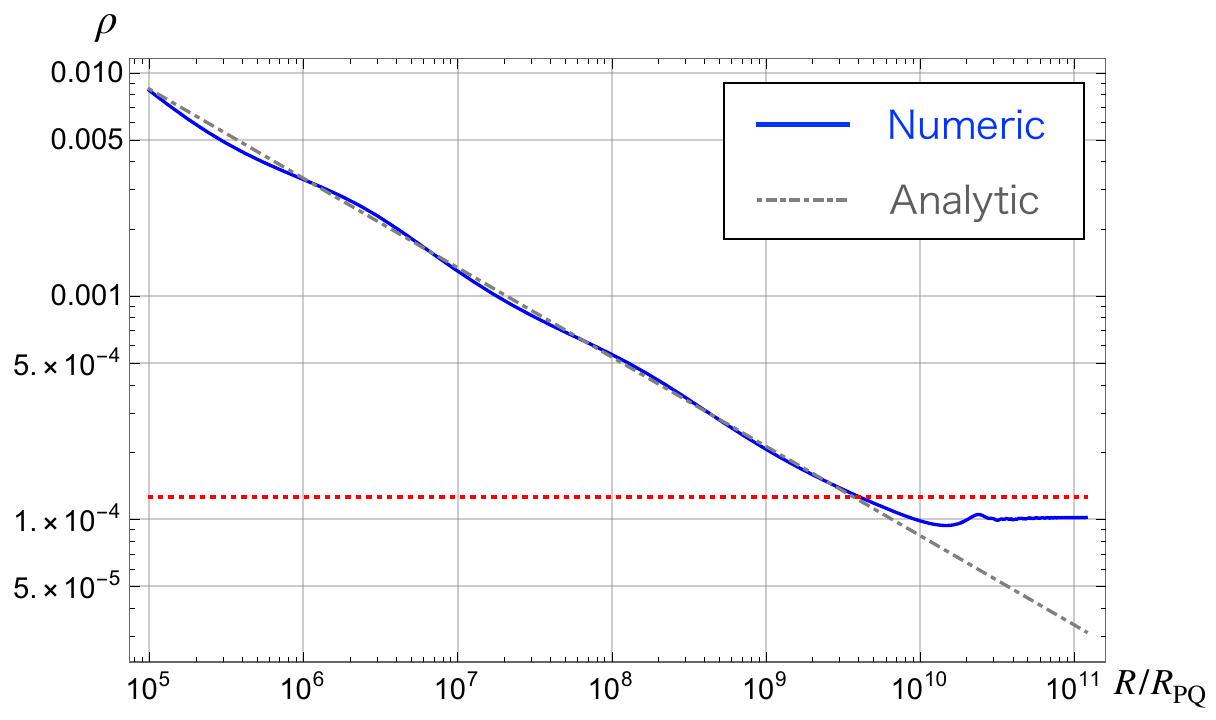}
\end{minipage}
\caption{
Evolution of fields, $\theta,\theta_b$, and $\rho$ as a function of the scale factor.
While the black solid and red dashed lines in the left panel represent the evolution of $\theta$ and $\theta_b$, the blue line in the right panel does that of $\rho$ with the scaling solution (\ref{SRD}) denoted by the gray dot-dashed line.
The red dotted line shows the field value of $\rho$ from which the backreaction becomes non-negligible.
Here we set $f_a=10^{10}\GeV$, $(N_{\rm DW},\ell,m,n)=(2,3,9,6)$, $|\lambda|=0.5$, $\lambda_S=10^{-4}$, $m_S=10^{-20}\GeV$, and $\delta=1$ for the initial conditions, $\theta_{\rm ini}=\pi/5$, $\theta_{b,{\rm ini}}=\pi/10$, $\chi_{\rm ini}$ given by \EQ{SRD}. 
}
\label{fig:evolution}
\end{figure*}

\subsection{Dynamics of the full system
\label{sec:full}}

Here let us discuss the dynamics of the full system to study the backreaction.
We note that the radial component of $P$ no longer affects the dynamics, as we assume that the Mexican-hat potential (the first term in \EQ{VPQ}) is dominant.
In the basis (\ref{parameterize}), the Lagrangian is given by
\beq
\Lag = \frac{1}{2}(\del_\mu a)^2 +\frac{1}{2}(\del_\mu \chi)^2 + \frac{1}{2}\chi^2(\del_\mu \theta_b)^2 -V_{\rm PQ} \ .
\eeq
Hereafter, let us define dimensionless fields as
\beq
\theta \equiv \frac{a}{v_{\rm PQ}} \ , \quad  \theta_b \equiv \frac{b}{\chi} \ , \quad \rho \equiv \frac{\chi}{\bar{\chi}} \ ,
\eeq
with a normalization factor,
\beq
\bar{\chi} = \left[\frac{2^{(m+\ell)/2-1}m!\ell!\Mpl^{m+\ell-4} H_{\rm PQ}^2}{|\lambda|v_{\rm PQ}^{\ell-2}}\right]^{1/m},
\eeq
where $H_{\rm PQ}$ is the Hubble parameter at $T=v_{\rm PQ}$.
Then we obtain the equations of motion,
\beq
&&\ddot{\theta} +3H\dot{\theta} + \ell H_{\rm PQ}^2 \rho^m \sin(\ell\theta + m\theta_b + \delta)\nonumber\\
&& ~~~~ +\frac{m_a^2f_a}{v_{\rm PQ}}\sin(N_{\rm DW}\theta) = 0 \ ,\\
&&\ddot{\theta}_b + 3H\dot{\theta}_b + 2\frac{\dot{\rho}}{\rho} \dot{\theta}_b\nonumber\\ 
&&~~~ + \frac{mv_{\rm PQ}^2H_{\rm PQ}^2}{\bar{\chi}^2} \rho^{m-2} \sin(\ell\theta + m\theta_b + \delta) = 0 \ ,\\
&&\ddot{\rho} + 3H\dot{\rho} + \frac{n\lambda_S^2}{2^{n-1}(n!)^2}\frac{\bar{\chi}^{2n-2}}{\Mpl^{2n-4}} \rho^{2n-1} -\dot{\theta}^2_b \rho + m_S^2 \rho\nonumber\\
&& ~~ -\frac{mv_{\rm PQ}^2H_{\rm PQ}^2}{\bar{\chi}^2} \rho^{m-1}\cos(\ell\theta + m\theta_b + \delta) = 0 \ .
\eeq

\FIG{fig:evolution} shows the time evolution of the fields as a function of the scale factor $R/ R_{\rm PQ}$ with $R_{\rm PQ}$ at $T=v_{\rm PQ}$.
Here we set $f_a=10^{10}\GeV$, $(N_{\rm DW},\ell,m,n)=(2,3,9,6)$, $|\lambda|=0.5$ (which is beyond \EQ{backreaction}), $\lambda_S=10^{-4}$, $m_S=10^{-20}\GeV$, $\delta=1$, and assume constant numbers of effective degrees of freedom for energy density and entropy density, $g_*=g_{*s}=106.75$.
We take the initial conditions $\theta_{\rm ini}=\pi/5$, $\theta_{b,{\rm ini}}=\pi/10$, and $\chi_{\rm ini}$ is taken in accordance with the scaling solution (\ref{SRD}) at $T=v_{\rm PQ}$.
First, the axion motion is driven by $V_{\cancel{\rm PQ}}$, but $\theta_b$ still remains static. 
When the potential from the QCD effect becomes dominant over $V_{\cancel{\rm PQ}}$, the axion oscillates around the minimum, inducing the small oscillation of $\theta_b$.
The behavior of $\rho$ looks very nontrivial and becomes constant when the backreaction cannot be neglected in the vicinity of the red dotted line derived from Eq.~(\ref{rhoconst}).
This is because $\theta$ and $\theta_b$ move to the minimum of $V_{\cancel{\rm PQ}}$, which makes the sign of the mixing term negative, and then the competition between $|S|^{2n}$ and $S^mP^\ell$ stabilizes a nonzero VEV of $S$.
Since the vacuum structure is not altered by the evolution of phases or mass term $m_S$, the PQ mechanism works well, even with a large amplitude of $|S|$. 
In the present paper, we focus on the simpler scenario where the extra potential disappears below the QCD scale by considering the bound (\ref{backreaction}), while the otherwise region will be discussed in \SEC{sec:Discussion}.

\begin{figure*}[t!]
\begin{minipage}[t]{16.5cm}
\centering
\includegraphics[width=8cm]{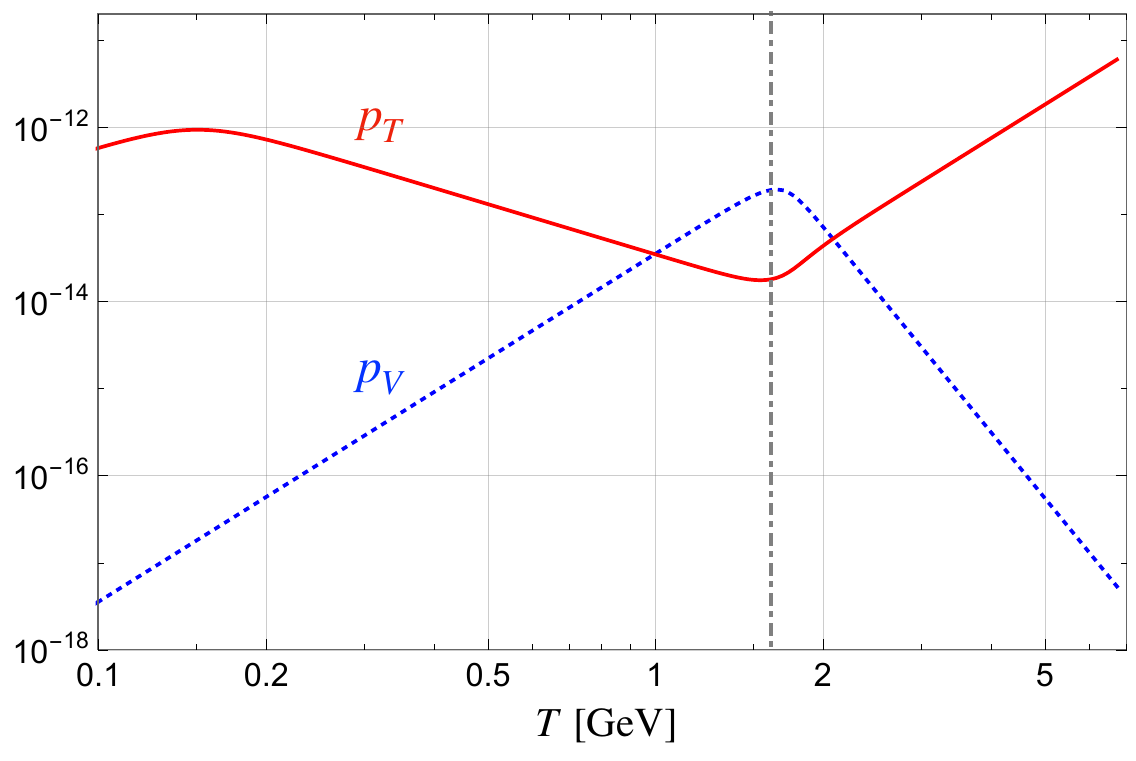}
\hspace{3mm}
\centering
\includegraphics[width=8cm]{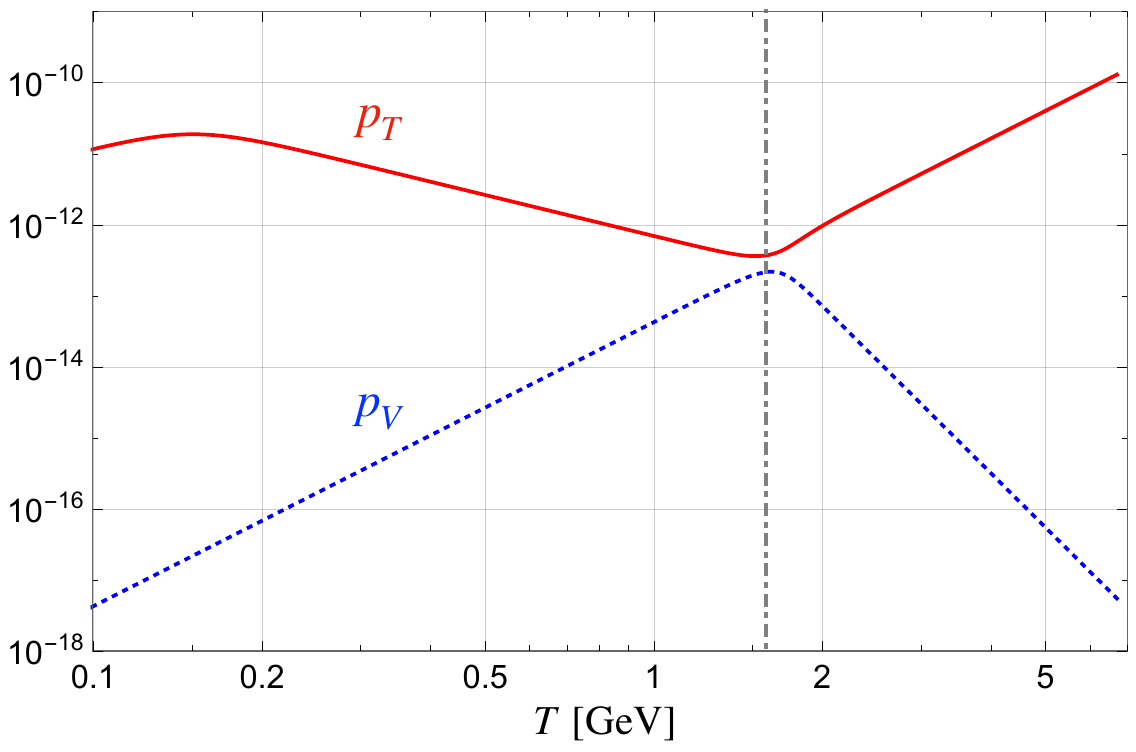}
\end{minipage}
\caption{
The tension force (red solid) and volume pressure (blue dotted) in units of $\GeV^4$ as a function of temperature.
We set $(N_{\rm DW}, \ell, m, n)=(3,2,10,6)$, $\lambda_S=10^{-4}$, $m_S=10^{-20}\GeV$, $f_a=10^{9}\GeV~(2\times10^{10}\GeV)$, and $\lambda=\lambda^{(\rm max)}\simeq 0.01~(3\times10^{-5})$ in the left (right) panel.
The gray vertical line denotes the temperature $T_{\rm tr}$ (\ref{Ttr}).
}
\label{fig:pTpV}
\end{figure*}

\section{Topological defects
\label{sec:defect}}

At $T\sim v_{\rm PQ}$, $P$ acquires a nonzero VEV, and the PQ symmetry is spontaneously broken.
When it occurs after inflation, cosmic string appears.
In our scenario where the axion starts to oscillate before the QCD crossover, domain walls are produced through the mixing-induced potential and connected with the cosmic strings, which is called the string-wall system.
The number of domain walls attached with a cosmic string is given by $\ell$.

While the system is unstable for $\ell=1$, it cannot collapse by itself for $\ell>1$ because the cosmic string is sustained by the tension of walls from multiple directions.
Here, we discuss the evolution and the fate of the string-wall system for each case.

\subsection{Short-lived system}

First let us consider the case of $\ell=1$, which was studied in ref.~\cite{Ibe:2019yew}.
Since one domain wall is attached with each cosmic string, when the tension force acting on the domain wall is comparable to that on the string, the system becomes unstable.
Given the wall and string tensions at $T \gg \Lambda_{\rm QCD}$,
\beq
\sigma_{\rm wall}(T) &\simeq& \frac{8}{\ell^2}m_{\cancel{\rm PQ}}(T)v_{\rm PQ}^2 \ ,  \\[1ex]
\mu_{\rm string}(T) &\simeq& 2\pi v_{\rm PQ}^2 \ln\lmk\frac{v_{\rm PQ}}{H(T)}\rmk  ,
\eeq
we can estimate the condition that the system is broken as \cite{Ibe:2019yew}
\beq
\frac{\sigma_{\rm wall}H}{\mu_{\rm string}H^2} &\gtrsim& 1\\
\leftrightarrow~
m_{\cancel{\rm PQ}}(T) &\gtrsim& \frac{\pi\ell^2}{4}H(T) \ln\lmk\frac{v_{\rm PQ}}{H(T)}\rmk,
\eeq
for $T<T_{\rm osc}$.
Naively, the condition is $m_{\cancel{\rm PQ}}\gtrsim \mathcal{O}(10)H$.
Even if $N_{\rm DW}>1$, the string-wall system via the potential from QCD effects no longer forms, because the axion field at any spacial point settles into the single minimum of $V_{\cancel{\rm PQ}}$.

\subsection{Long-lived system}

For $\ell>1$, the string-wall system must be broken due to some destabilization effects.
If the string-wall system is never broken or too long-lived, the domination of energy density or the overproduced axions would spoil the success of the standard cosmology, but once it collapses successfully, the axion DM can be produced with the correct abundance.
From the perspective of the system made up by $V_{\cancel{\rm PQ}}$, the QCD potential seems to bias the symmetric structure, as long as $\ell$ and $N_{\rm DW}$ are co-prime number (see ref.~\cite{Kitajima:2023cek,Cyr:2025nzf,Notari:2025kqq} for time-dependent bias potential terms). 
However, we will find it nontrivial to conclude the fate of the system.

Let us begin with the discussion on the stability of the system in (semi)analytical way.
There are two kinds of destabilization effects: one is a biased potential, the other is some structural instability.
First, we focus on the biased potential.\footnote{
Although both effects should be taken into account simultaneously, we take a limit that either effect is dominant in the following discussion.}
A bias term against the string-wall system generates the volume pressure, which is estimated by the potential difference between two vacua, $p_V\simeq \Delta V_{\rm bias}$.
When the volume pressure becomes comparable to the tension force $p_T$, the system starts to be destabilized.
In our case, while $V_{\rm bias} = V_{\rm QCD}$ at $T\gg\Lambda_{\rm QCD}$, $V_{\rm bias} = V_{\cancel{\rm PQ}}$ at $T\ll\Lambda_{\rm QCD}$, and the domain wall number can change from $\ell$ to $N_{\rm DW}$.
To validate the system collapse in this complicated setup, we can numerically estimate $p_V$ and $p_T$.
The tension force is given by
\beq
p_T \simeq 4H\int_{\theta_{{\rm min}1}}^{\theta_{{\rm min}2}} d\theta \frac{|V(\theta)|}{\sqrt{m_{\cancel{\rm PQ}}^2+m_a^2}} \ ,
\eeq
where $V=V_{\cancel{\rm PQ}}+V_{\rm QCD}$, and $\theta_{{\rm min}1},\theta_{{\rm min}2} \, (>\theta_{{\rm min}1})$ are the adjacent local minima of $V_{\cancel{\rm PQ}}$.
The prefactor is chosen, so that $p_T\sim 2\sigma_{\rm wall}H$ in the high temperature limit.\footnote{Some numerical simulations indicate the deviation from the standard scaling law, $\rho_{\rm wall}~({\rm or}~p_T)\propto R^{-3}$, which might reduce the final abundance of the axion and gravitational waves \cite{Babichev:2025stm}.}
\FIG{fig:pTpV} describes the temperature dependence of the tension force $p_T$ (red solid) and the volume pressure $p_V$ (blue dotted) for $(N_{\rm DW}, \ell, m, n)=(3,2,10,6)$, $\lambda_S=10^{-4}$, $m_S=10^{-20}\GeV$, $f_a=10^{9}\GeV~(2\times10^{10}\GeV)$, and $\lambda=\lambda^{(\rm max)}\simeq 0.01~(3\times10^{-5})$ in the left (right) panel.
We take the maximum value of $\lambda$ to satisfy the backreaction constraint.
The vertical gray line is estimated by $|V_{\rm QCD}|\sim |V_{\cancel{\rm PQ}}|$, or 
\beq
&&m_{\cancel{\rm PQ}}(T_{\rm tr}) \sim \frac{\ell}{N_{\rm DW}} m_a(T_{\rm tr})\\
&\leftrightarrow& T_{\rm tr} \simeq 1.6\GeV \lmk\frac{|\lambda|}{0.01}\rmk^{-\alpha} \lmk\frac{v_{PQ}}{2\times 10^{9}\GeV}\rmk^{-\ell\alpha},\nonumber\\
\label{Ttr}
\eeq 
where $\alpha\equiv (2\tilde{b}+2m/(n-1))^{-1}$.
At the boundary $T=T_{\rm tr}$, the role of each potential is switched from bias to tension or vice versa.
Assuming that it takes $\Delta T\sim \mathcal{O}(1)\GeV$ conservatively for the system to collapse completely, we find that $f_a\lesssim10^9\GeV$ is required, as shown in the left panel of \FIG{fig:pTpV}.
This assumption is naively justified by the results of simulations
\cite{Kawasaki:2014sqa}.

\begin{figure}[t!]
\centering
\includegraphics[width=5cm]{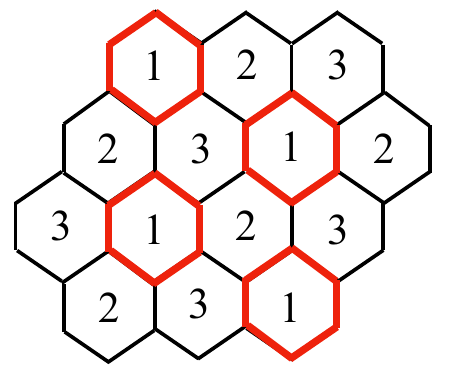}
\caption{
Approximate description of the system for $\ell=3$ and $N_{\rm DW}=2$ (a 2D slice perpendicular to the strings is shown).
After $V_{\rm QCD}$ becomes dominant, the vacua numbered by 2 and 3 are integrated, and one vacuum is isolated.
The black lines denote the initial domain walls, while the red lines the final domain walls.
}
\label{fig:large_l}
\end{figure}

Even for $f_a\gtrsim10^{9}\GeV$, the system is able to be destabilized by the structural instability.
The mixing-induced potential makes the initial distribution of the axion field non-uniform.
When the axion starts to oscillate by $V_{\cancel{\rm PQ}}$, the axion field configuration is given by the domain wall solution, which is highly inhomogeneous in space.
When $V_{\rm QCD}$ is turned on, the local minima of the potential are deformed and a new structure combining domain walls and strings appears around $T\simeq T_{\rm tr}$, but in some case, the system cannot be stabilized in a symmetric way. Let us consider a particular case when $\ell>N_{\rm DW}$ such as $\ell=3$, $N_{\rm DW}=2$ (see \FIG{fig:large_l}).
Initially, three domain walls are attached to one string in the most symmetric way.
After $V_{\rm QCD}$ becomes dominant, one vacuum disappears and the vacuum structure is decomposed into the remaining two vacua. 
Due to the initial configuration, we can expect that one of them is isolated and collapses by the tension.

It should be noted that the system may not be destabilized by an asymmetric structure for $m\geq2n-1$. 
Since $m_{\cancel{\rm PQ}}$ decreases faster than $H$ and becomes smaller than $H$ at some temperature, the axion configuration can be rearranged by the gradient term.
For a larger value of $\lambda$, the re-randomization might be mild, but we conservatively focus on $m\leq2n-2$ in the following discussion.

So far, we have qualitatively discussed the destabilization of the system only in one specific case.
For general combinations of $(\ell, N_{\rm DW})$, the situation is less clear.
While we comment on this issue in \SEC{sec:Discussion}, a detailed analysis is left for future work. Here, we present conservative conditions for destabilizing the system:
\beq
f_a\lesssim10^{9}\GeV \ , \quad \text{for} \ ^{\forall} (\ell, N_{\rm DW}) \ ,
\eeq
or
\beq
(\ell, N_{\rm DW}) = (3,2) \ , \quad \text{for} \  ^{\forall}f_a~{\rm and}~m\leq2n-2 \ .
\eeq
Note that $\forall$ means all possible combinations or values of $f_a$ which are restricted by the backreaction constraints.
Moreover, there remain curious questions for the latter case, how much time it takes to collapse or how the dynamics depends on the size of $V_{\cancel{\rm PQ}}$.
To answer those questions requires a detailed simulation, which is left for a future study.
Here we simply assume that the annihilation occurs after the potential minima are deformed, i.e. at $T_{\rm ann} = \kappa T_{\rm tr}$, with $\kappa < 1$ a numerical parameter.
Since the hexagons can shrink with the shape kept, the typical time scale is $H^{-1}$, or $\kappa\sim\mathcal{O}(0.1)$ can be taken.
The latter case is severer for the domain wall problem, and we focus on this case below.

Let us now estimate the abundance of the axion produced from the domain wall annihilation.
Since the energy density of strings is diluted faster than that of domain walls, we here ignore the dynamics of strings.
The evolution of energy densities of domain walls, axions, and gravitational waves are respectively given by
\beq
\frac{d\rho_{\rm wall}}{dt} &=& -(1+p)H\rho_{\rm wall} -\left.\frac{d\rho_{\rm wall}}{dt}\right|_{\rm emission}
\label{wallEQ} \ , \\[1ex]
\frac{d\rho_a}{dt} &=& -3H\rho_a + \frac{d\rho_{w\rightarrow a}}{dt} \ , \\[1ex]
\frac{d\rho_{\rm gw}}{dt} &=& -(4+2p)H\rho_{\rm gw} + \frac{d\rho_{w\rightarrow gw}}{dt} \ ,
\label{gwEQ}
\eeq
where $p=m/(n-1)$, and we assume that the total amount of the decay products from the domain wall consists of the axion and gravitational waves,
\beq
\left.\frac{d\rho_{\rm wall}}{dt}\right|_{\rm emission}
= \frac{d\rho_{w\rightarrow a}}{dt} + \frac{d\rho_{w\rightarrow gw}}{dt} \ .
\eeq
We assume the system follows the scaling law,
and the domain wall tension is temperature dependent, or $\rho_{\rm wall}\propto H^{m/2(n-1)+1}\propto R^{-m/(n-1)-2}$.
In the scaling regime, the energy density of gravitational waves is given by $\rho_{\rm gw} \simeq \epsilon_{\rm gw} G \sigma_{\rm wall}^2$, where $\epsilon_{\rm gw}$ is the efficiency of production.
The dilution by the scaling law is compensated by the second terms in Eqs.~(\ref{wallEQ}), (\ref{gwEQ}).
We then obtain the rate of the axion production,
\beq
\frac{d\rho_{w\rightarrow a}}{dt} \simeq \frac{\sigma_{\rm wall}}{2t^2} -2\epsilon_{\rm gw}\frac{G\sigma_{\rm wall}^2}{t} \ .
\eeq
As the potential from the QCD effects grows up, the tension force or the energy density of the system evolves as shown in \FIG{fig:pTpV}.
Assuming that the produced axion has energy $\sqrt{1+\epsilon_a^2}m_a$ where $\epsilon_a$ is defined as the momentum of the produced axion in units of $m_a$, the comoving number density of the axion is given by
\beq
&&N_{a,{\rm dec}}(t) \equiv R^3(t) n_{a,{\rm dec}}(T)\nonumber\\[1ex]
&\simeq& \int_{t_{\rm osc}}^t dt' \frac{R^3(t')}{\sqrt{1+\epsilon_a^2}m_a(t')} \frac{d\rho_{w\rightarrow a}}{dt'}\nonumber\\[1ex]
&\simeq& \frac{R^3(t)}{\sqrt{1+\epsilon_a^2}m_a(t)} \left[\frac{\sigma_{\rm wall}}{t}-\frac{4}{3-p}\epsilon_{\rm gw}G\sigma_{\rm wall}^2(t)\right].\nonumber\\
\label{Nadec}
\eeq
Here $n_{a,{\rm dec}}$ denotes the number density of the produced axion, and we have used $t\gg t_{\rm osc}$ in the third equality.
We can assume $\sigma_{\rm wall} \simeq 8m_a f_a^2$.
In addition, the string decay also produces the axion from $T\simeq v_{\rm PQ}$ to $T=T_{\rm osc}$.
Since we focus on $T_{\rm osc}\gg T_{\rm osc}^{(\rm conv)}$, the string decay stops at a very early epoch, leading to the only subdominant contribution.

By using \EQ{Ttr} and \EQ{Nadec}, we obtain the abundance of the axion produced from the domain wall,
\beq 
&&\Omega_{a,{\rm dec}}h^2 = \frac{m_{a,0}N_{a,{\rm dec}}(t_{\rm ann})}{\rho_{\rm crit}h^{-2}R_0^3}\nonumber\\
&\simeq& 0.12 \frac{1}{\sqrt{1+\epsilon_a^2}} \lmk\frac{\kappa}{0.1}\rmk^{-1} \lmk\frac{|\lambda|}{2\times10^{-4}}\rmk^{\alpha}\nonumber\\
&\times& \lmk\frac{N_{\rm DW}}{2}\rmk^{\ell\alpha} \lmk\frac{f_a}{2.4\times10^{10}\GeV}\rmk^{1+\ell\alpha},
\label{Omegadec}
\eeq
where we set $(\ell,m,n)=(3,9,6)$ and $\lambda_S=10^{-4}$, and the contribution of gravitational waves is negligibly small.
Since the annihilation temperature can be lower than that used in the estimation and $N_{a,{\rm dec}}\propto T_{\rm ann}^{-1}$, we note that the abundance can be enhanced, so that the upper bound on $f_a$ would be severer.
Although a smaller $m$ can be also taken, the upper bound on $f_a$ from the backreaction and $T_{\rm osc}>T_{\rm osc}^{(\rm conv)}$ become tighter.
Thus the domain wall problem is solved, as long as it is solved for $(\ell,m,n)=(3,9,6)$.

\section{Axion DM from misalignment
\label{sec:misalignment}}

\begin{figure*}[t!]
\begin{minipage}[t]{16.5cm}
\centering
\includegraphics[width=7.9cm]{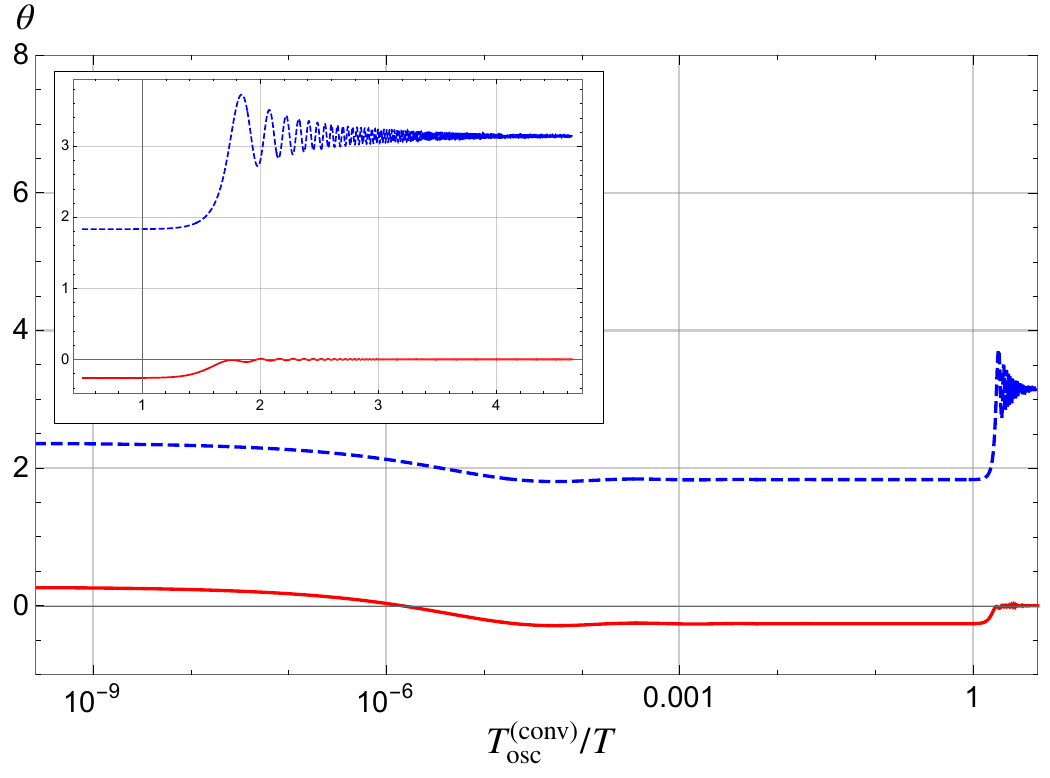}
\hspace{2mm}
\centering
\includegraphics[width=8.1cm]{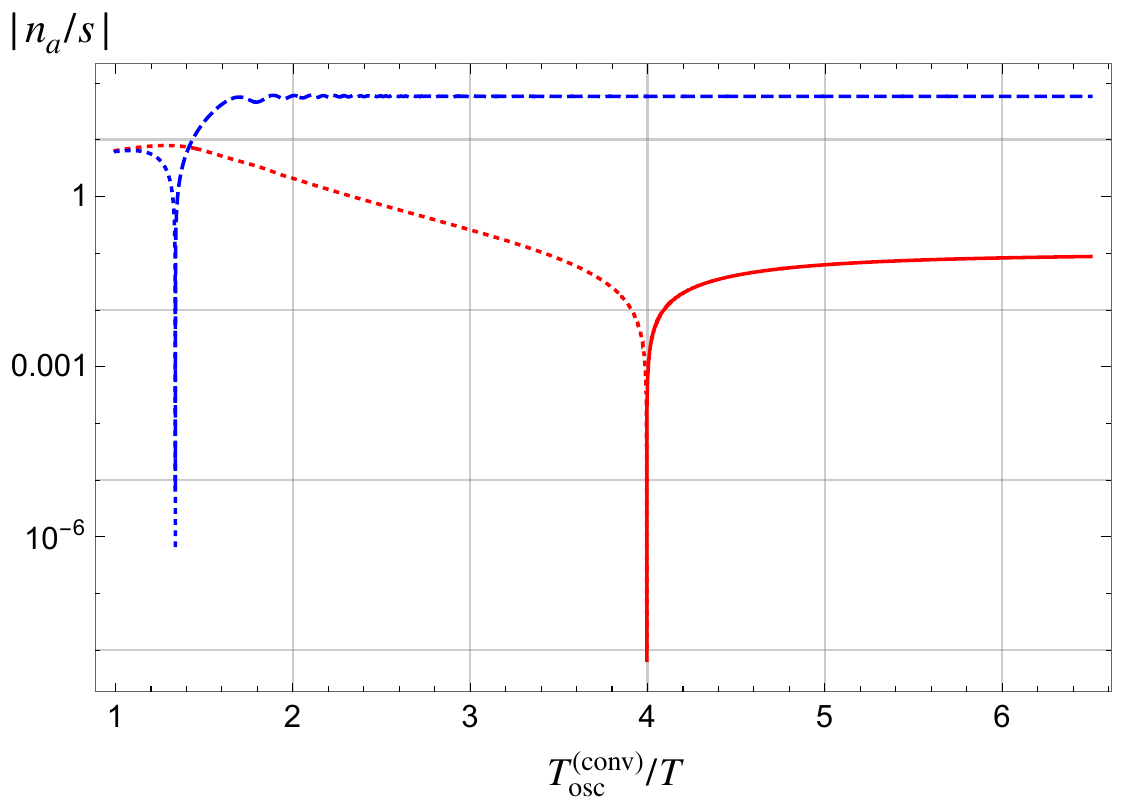}
\end{minipage}
\caption{
Evolution of the axion field (left) and the number density (right) as a function of $T_{\rm osc}^{(\rm conv)}/T$.
The region of $T_{\rm osc}^{(\rm conv)}/T = \mathcal{O}(1)$ is enlarged in the window inside the left panel.
We set $f_a=10^{10}\GeV$, $(N_{\rm DW},\ell,m,n)=(2,3,9,6)$, $|\lambda|=3\times10^{-3}$, $\lambda_S=10^{-4}$, and $\delta'=\pi/4$.
The red solid and blue dashed lines represent the evolution for the smooth shift regime and the trapped regime, respectively.
In the right panel, we represent the negative value by the dotted lines.
The initial oscillation amplitude of the axion field is taken to be $\pi/6$.
}
\label{fig:axion}
\end{figure*}

We here consider the production of the axion DM via the (trapped) misalignment mechanism.
The isocurvature perturbation constraint is also discussed.
Finally, we find the total axion DM abundance from the domain wall decay and the misalignment mechanism,
and clarify the viable parameter space to address the domain wall problem.

\subsection{Axion field dynamics}

In addition to topological defects, coherent oscillations of the axion field contribute to the DM abundance via the misalignment mechanism \cite{Preskill:1982cy,Abbott:1982af,Dine:1982ah}.
The extra potential induced by the mixing term makes the axion evolution more nontrivial, which can be classified into the so-called trapped misalignment \cite{Kawasaki:2015lpf,Higaki:2016yqk,Kawasaki:2017xwt,Nakagawa:2020zjr,DiLuzio:2021pxd,DiLuzio:2021gos,Jeong:2022kdr,Nakagawa:2022wwm,DiLuzio:2024fyt}.
The dynamics is classified into two regimes
\cite{Jeong:2022kdr}:
\beq
&{(\rm I)}& ~~~|\theta_{\rm min}^{(\cancel{\rm PQ})}-\theta_{\rm min}^{(\rm QCD)}|\ll\frac{\pi}{\ell}~~~({\rm smooth~shift~regime}) \ , \nonumber\\[1ex]
&{(\rm II)}& ~~~|\theta_{\rm min}^{(\cancel{\rm PQ})}-\theta_{\rm min}^{(\rm QCD)}|\gtrsim\frac{\pi}{\ell}~~~({\rm trapped~regime}) \ .\nonumber
\eeq
Here we define $\theta_{\rm min}^{(\cancel{\rm PQ})}\equiv (2\pi k-\delta')/\ell$ as the minimum of $V_{\cancel{\rm PQ}}$ at which the axion first oscillates, with $\delta'\equiv \delta+m\theta_b$ (which can be identified as a free parameter) and $k\in \mathbb{Z}$, and $\theta_{\rm min}^{(\rm QCD)}$ is taken as the closest minimum of $V_{\rm QCD}$ to it.
In the smooth shift regime, if the trapping effect of the extra potential is strong enough, any additional oscillation is not induced when the potential minimum shifts due to $V_{\rm QCD}$.
This means that the axion follows the shift of the potential minimum, and the suppression occurs when $|\dot{m}_{\rm eff}/m_{\rm eff}^2|\ll1$ where $m_{\rm eff}\equiv\sqrt{|V''(a)|}$ evaluated at the temporal minimum \cite{Nakagawa:2020zjr}.
This is called the adiabatic suppression mechanism \cite{Linde:1996cx} (\cite{Kawasaki:2015lpf,Kawasaki:2017xwt,Nakagawa:2020zjr} for the QCD axion).
In the trapped regime, the axion is trapped at the wrong vacuum for a longer time, and after the wrong vacuum vanishes, the axion starts to oscillate again around the CP conserving minimum.

To recognize the two regimes, we show the time evolution of the axion field and the absolute value of the number density normalized by the entropy density, $s$, as a function of $T_{\rm osc}^{(\rm conv)}/T$ in \FIG{fig:axion}.
Here we choose an example parameter set $f_a=10^{10}\GeV$, $(N_{\rm DW},\ell,m,n)=(2,3,9,6)$, $|\lambda|=3\times10^{-3}$, $\lambda_S=10^{-4}$, $\delta'=\pi/4$, and use the temperature-dependent $g_*$ and $g_{*s}$ estimated in ref.~\cite{Saikawa:2018rcs}.
The value of $|\lambda|$ corresponds to the maximum value (\ref{backreaction}) for $m_S\sim10^{-20}\GeV$. 
The red solid and blue dashed lines represent the evolutions for the smooth shift regime and the trapped regime, respectively. 
In the right panel, we represent the negative value by the dotted lines.
The initial conditions are chosen, so that the initial oscillation amplitude is set to $|\theta_{\rm ini}-\theta_{\rm min}^{(\cancel{\rm PQ})}| = \pi/6$.\footnote{
The negative value of $n_a/s$ in Fig.~\ref{fig:axion} arises from the choice of vacuum energy normalization in the presence of a time-dependent $V_{\cancel{\rm PQ}}$. In our current normalization, $\rho_a$ initially becomes negative. Once the spectator field $S$ starts oscillating, $n_a/s$ becomes well-defined as a number density. We confirm the approximate number conservation in Fig.~\ref{fig:axion}. 
}
One can see that the number density in the case (I) is much smaller than that in the case (II).
This is not mainly because the (second) oscillation amplitude is small, but because the amplitude is slightly more suppressed as well as the delayed oscillation for the case (II).
However, the oscillation is not strongly suppressed, because the backreaction constraint put the upper bound on $|\lambda|$ (\ref{backreaction}) and the trapping effect cannot be very strong in our model.
To clarify the difference from the conventional smooth shift regime, we hereafter rename it as `{\it pseudo-smooth shift regime}', which does not show sufficient adiabaticity.
Such an intermediate range of the amplitude of $V_{\cancel{\rm PQ}}$ has not been studied so far.
Thus, we perform a numeric estimate of the axion abundance in addition to supports with analytical method.

\subsection{Abundance}

In the post-inflationary scenario, the abundance of the axion DM can be estimated as the average of contributions from all possible initial values of the axion field.
The number density is given by
\beq 
n_a \simeq \frac{1}{2\pi}\int_{-\pi}^{\pi} d\theta_{\rm ini}  n_a(\theta_{\rm ini}) \ ,
\eeq
where $n_a(\theta_{\rm ini})$ represents the number density for an initial value of the axion field $\theta_{\rm ini}$.

To perform a numerical estimation, we use several approximations.
First, we assume that when the axion field oscillates around a common minimum $\theta_{\rm min}^{(\cancel{\rm PQ})}$, the abundance does not depend on $\theta_{\rm ini}$.
This assumption is justified unless $V_{\cancel{\rm PQ}}$ aligns with $V_{\rm QCD}$, i.e. $\left|\theta_{\rm min}^{(\cancel{\rm PQ})}-\theta_{\rm max}^{(\rm QCD)}\right|\ll1$.\footnote{
A $\theta_{\rm ini}$ dependence can arise even away from the alignment case due to anharmonic effects.
However, it is restricted to a very narrow range of $\theta_{\rm ini}$ and ignored in our estimation.
One can see it from \FIG{fig:thini} in Appendix \ref{app:adiabatic}.
}
As the second approximation, we use the formula (\ref{SRD}) for the scaling solution without following the full system.
Thirdly, the energy density is estimated as the time average of the kinetic energy,
\beq
\rho_a &=& \frac{\dot{a}^2}{2} + V_{\rm QCD}+V_{\cancel{\rm PQ}}\nonumber\\
&\simeq& 2\frac{1}{\delta t}\int_{\delta t}dt \frac{\dot{a}^2}{2} \ ,
\label{rhoa}
\eeq
after the kinetic energy amplitude normalized by the entropy density becomes constant.
Here $\delta t$ is taken to be much larger than $m_{a}^{-1}$.
Since the spectator field has a large amplitude until it oscillates when $m_S\sim H$, the axion has a non-negligible, time-dependent vacuum energy. 
To estimate the current abundance, we would need to analyze the full equations of motion at least just after $m_S\sim H$.
However, since the amplitude of the kinetic energy becomes constant after the oscillation around $V_{\rm QCD}$, which is finally expected to be comparable to the potential energy, it is sufficient to follow the axion equation of motion only until the oscillation around $V_{\rm QCD}$, reducing the computation time.

The assumptions in the numerical calculation are discussed in Appendix \ref{app:adiabatic} in detail.
In particular, we clarify the range of $|\lambda|$ where our assumptions are justified, and derive the analytical formula of the abundance.

We can have two types of estimation: (A) $\ell<N_{\rm DW}$ and (B) $\ell>N_{\rm DW}$.\footnote{As we mentioned in \SEC{sec:defect}, it is unclear if the string-wall system can collapse except for $(\ell,m,n)=(3,9,6)$.
However, we take several combinations in this section to show how the misalignment production is modified by the extra potential.}
In the following estimation, we focus on $m=9~(=2n-3)$ case, but we numerically confirmed that the overall behavior is the same as the case of $m=10~(=2n-2)$.

\begin{figure*}[t!]
\begin{minipage}[t]{16.5cm}
\centering
\includegraphics[width=8.0cm]{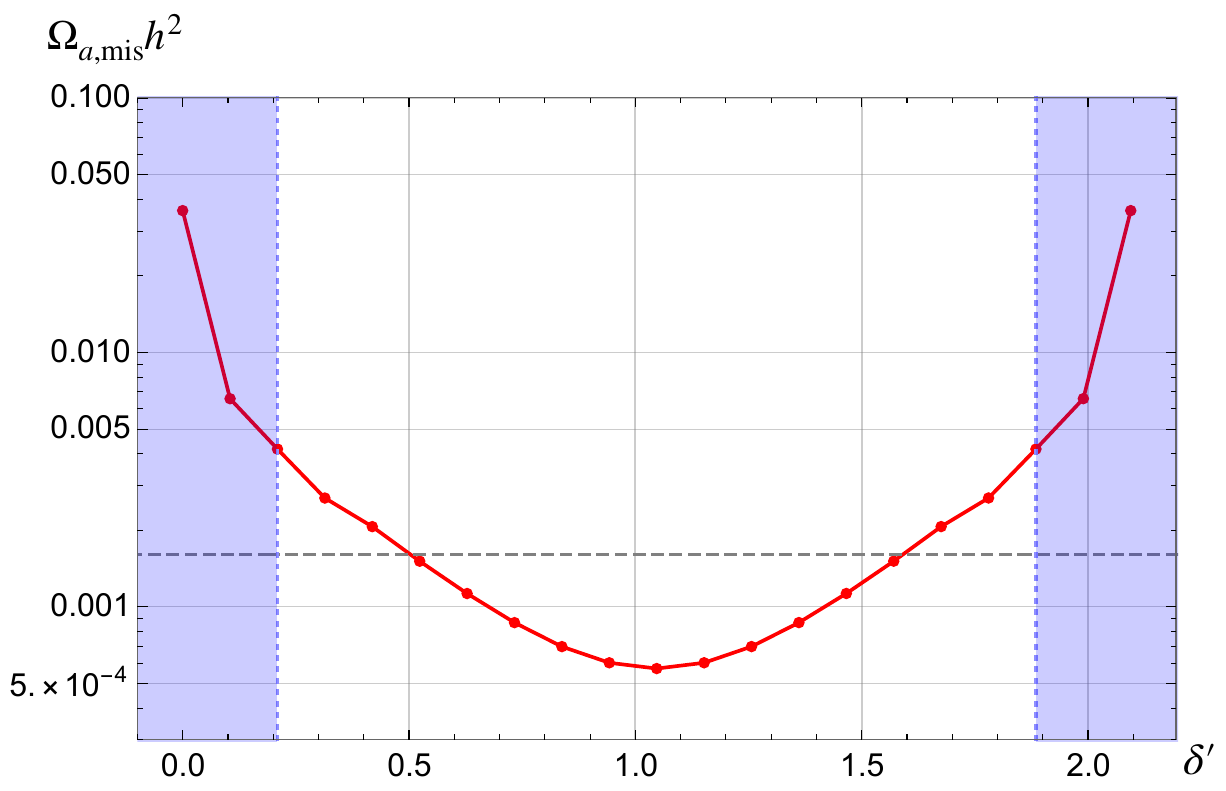}
\hspace{2mm}
\centering
\includegraphics[width=8.0cm]{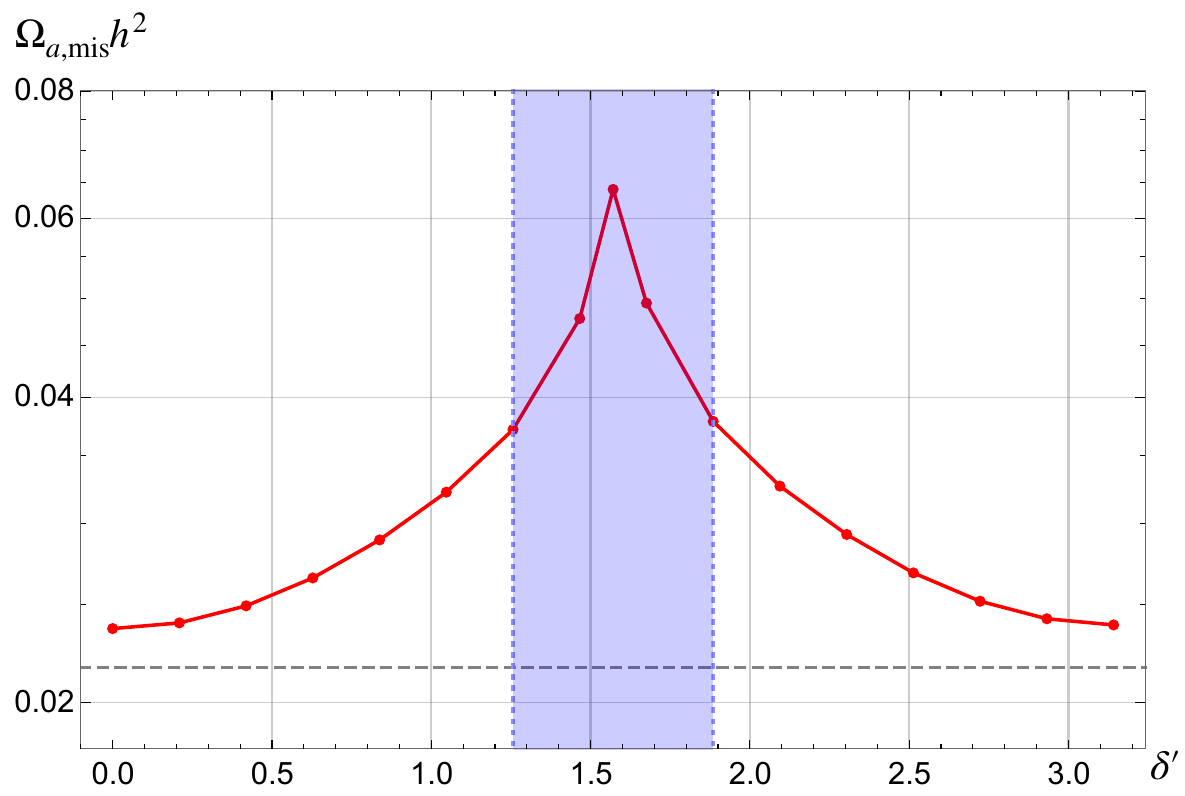}
\end{minipage}
\caption{
The abundance produced via the misalignment mechanism as a function of $\delta'$ for $N_{\rm DW}=3,~2$ in the left and right panels, respectively. 
While the left panel shows the result for $(\ell,m,n)=(2,9,6)$ and $f_a=10^{10}\GeV$,
the right panel shows for $(\ell,m,n) = (3,9,6)$ and $f_a=8\times10^{10}\GeV$.
In both panels, we set $|\lambda|=|\lambda^{(\rm max)}|$.
The gray dashed line denotes the abundance of the conventional QCD axion (\ref{convOmega}).
In the blue shaded bands, we show schematically the region where the approximation in our numerical calculation is broken.
}
\label{fig:abund}
\end{figure*}

\subsubsection{Case (A)}

In this case, there is no trapped regime, and the abundance can be basically suppressed by the adiabatic suppression mechanism. 
Assuming that the initial axion field is uniformly distributed, the number density of the axion is given by
\beq
n_a \simeq \frac{1}{\ell} \sum_{r=1}^{\ell} n_a^{(r)} \ .
\label{small_l}
\eeq
Here $n_a^{(r)}\equiv\rho_a^{(r)}/m_a$ is estimated for the case where the axion starts to oscillate around each $\theta_{\rm min}^{(\cancel{\rm PQ})}$.

In the left panel of \FIG{fig:abund}, we show the numerical result for $(N_{\rm DW},\ell,m,n)=(3,2,9,6)$ as a function of the relative phase $\delta'$. 
Here we choose an example parameter set, $f_a=10^{10}\GeV$, $\lambda_S=10^{-4}$, and $|\lambda|=|\lambda^{(\rm max)}|$.
The gray horizontal line denotes the conventional QCD axion abundance
\cite{Lyth:1991ub},
\beq
\Omega_{a,{\rm mis}}^{(\rm conv)}h^2 \simeq 1.6\times10^{-3} \lmk\frac{g_*}{80}\rmk^{\frac{\tilde{b}+2}{2(\tilde{b}+4)}}\lmk\frac{f_a}{10^{10}\GeV}\rmk^{\frac{\tilde{b}+6}{\tilde{b}+4}} \!\!\!\!\!\! .
\label{convOmega}
\eeq
Here the anharmonic effect in terms of $\theta_{\rm ini}$ is taken into account.
One can see that the abundance is slightly suppressed by the trapping effect.
At $\delta'=2\pi j/3$ with $j=0,1,2,\cdots$, where $V_{\cancel{\rm PQ}}$ aligns with $V_{\rm QCD}$, the abundance enhances because the axion evolution is more non-adiabatic due to the anharmonic effect \cite{Nakagawa:2020zjr}.
Our approximation gets worse in this range, because the dependence on $\theta_{\rm ini}$ is no longer negligible.
Requiring such a significant fine tuning does not motivate us to further explore this case.

\subsubsection{Case (B)}

For $\ell>N_{\rm DW}$, there are $(\ell-N_{\rm DW})$ trapped regimes.
Assuming the flat distribution of the initial axion field, we obtain the number density of the axion,
\beq
n_a \simeq \frac{\ell-N_{\rm DW}}{\ell} n_a^{(\rm trapped)} + \frac{N_{\rm DW}}{\ell} n_a^{(\rm smooth)} \ ,
\label{large_l}
\eeq
where $n_a^{(\rm trapped)}$ and $n_a^{(\rm smooth)}$ are the number densities for the trapped and pseudo-smooth shift regimes, respectively. 
The first contribution is dominant according to the numerical results in \FIG{fig:axion}.

In the right panel of \FIG{fig:abund}, the numerical result of the abundance for $(N_{\rm DW},\ell,m,n)=(2,3,9,6)$ are shown for $f_a=8\times10^{10}\GeV$, $\lambda_S=10^{-4}$, and $|\lambda|=|\lambda^{(\rm max)}|\simeq10^{-5}$.
We can see that the abundance is enhanced compared to that of the conventional QCD axion.
We note again that our approximation gets worse in the range of $|\delta'-2\pi(j+1/4) |\ll1$.

\begin{figure*}[t!]
\begin{minipage}[t]{16.6cm}
\centering
\includegraphics[width=8.2cm]{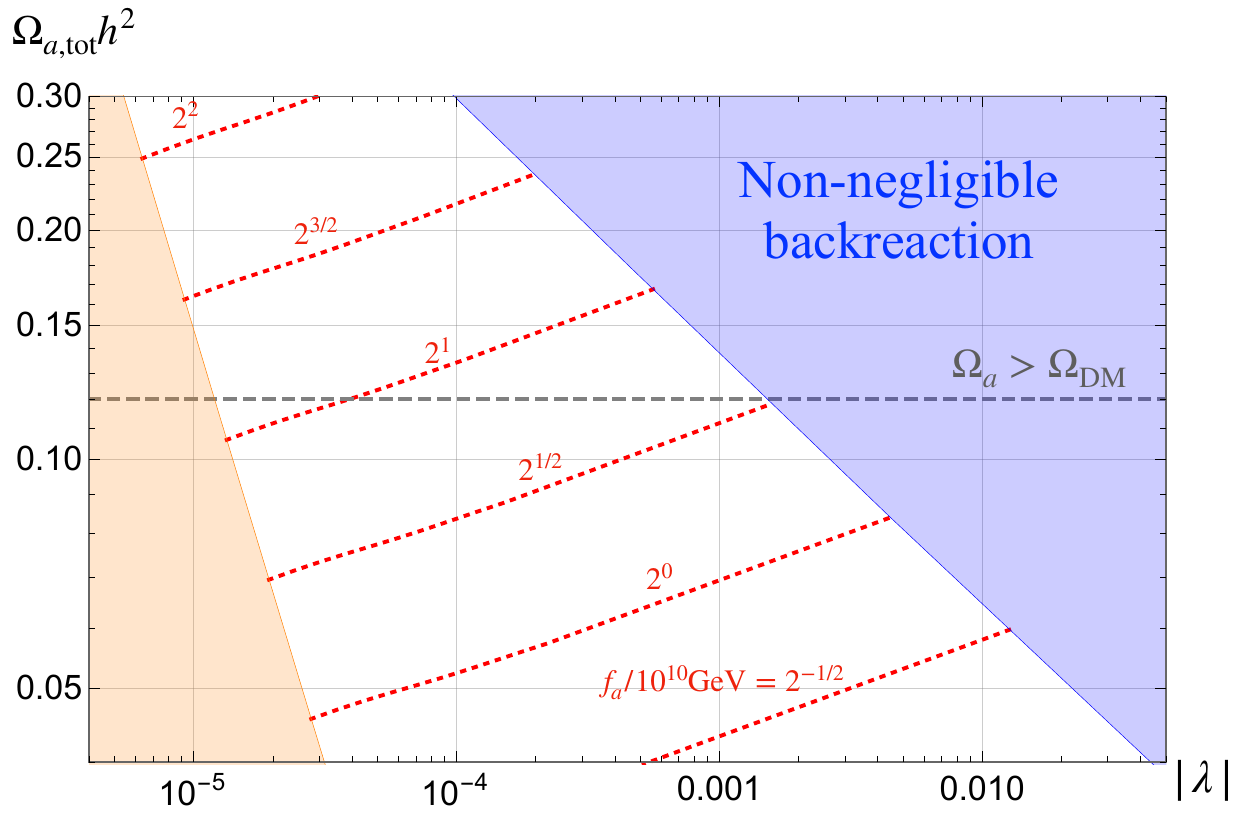}
\hspace{2mm}
\centering
\includegraphics[width=8.0cm]{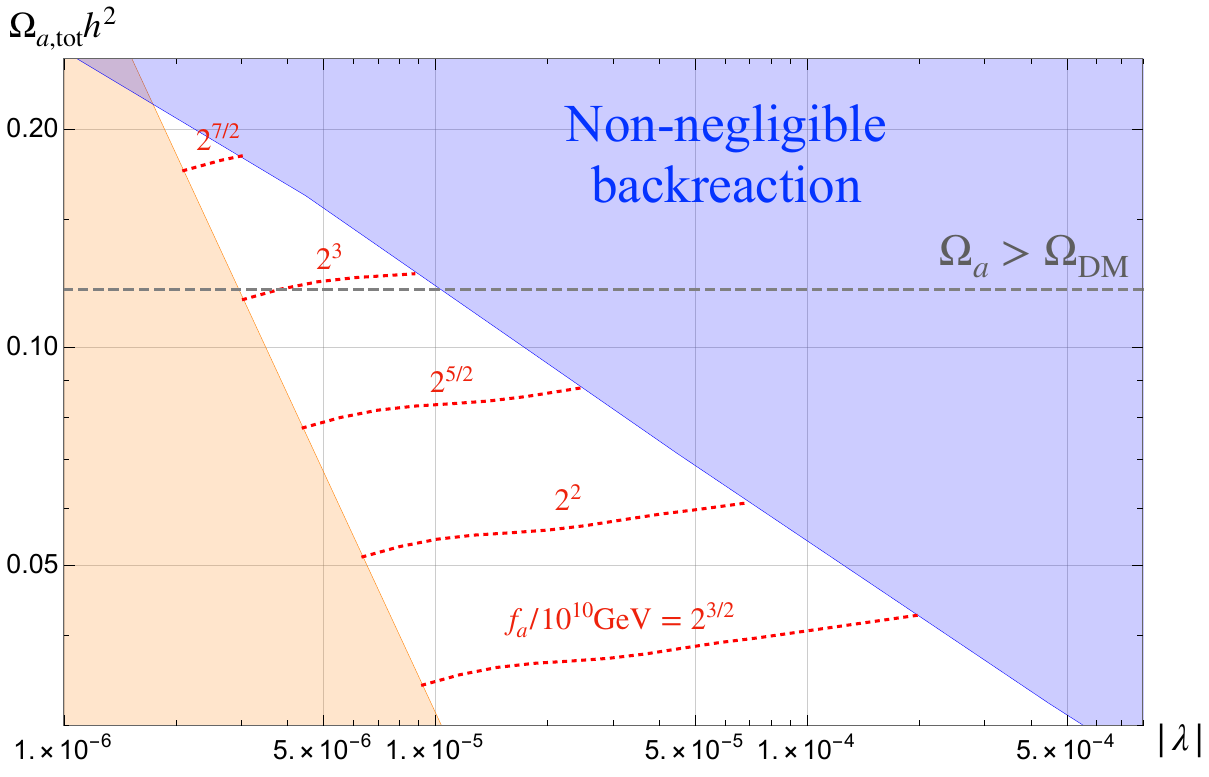}
\end{minipage}
\caption{
The total axion abundance $\Omega_{a,{\rm tot}}h^2$ as a function of $|\lambda|$ for $\delta'=1$, $\lambda_S=10^{-4}$,  $m_S=10^{-20}\GeV$, and $(N_{\rm DW},\ell,m,n)=(2,3,9,6)$.
We take $\kappa=0.1,0.5$ in the left and right panels, respectively.
The blue and orange shaded regions are excluded by the constraints of the backreaction (\ref{backreaction}) and the stability of the string-wall system.
The gray dashed line gives the correct DM abundance. 
}
\label{fig:viable}
\end{figure*}

\subsection{Isocurvature perturbations}

Let us briefly discuss the isocurvature perturbations of the axion field.
Although we assume that the PQ symmetry is spontaneously broken after inflation, the axion field has isocurvature fluctuations due to the mixing with the phase $b$ of $S$ which acquires quantum fluctuations during inflation \cite{Ibe:2019yew}.
The fluctuation of the phase $b$ during inflation is given by
\beq
\delta b_{\rm inf} \simeq \frac{H_{\rm inf}}{2\pi} \ .
\eeq
As $S$ obeys the scaling solution, the fluctuation evolves with time,
\beq
\delta b \simeq \frac{\chi}{\langle\chi_{\rm inf}\rangle} \delta b_{\rm inf} \ .
\eeq
Since $\delta b$ induces the fluctuations of center of the oscillation, the fluctuation of the axion field is estimated as 
\beq
\frac{\delta a}{f_a} \simeq \frac{mN_{\rm DW}}{\ell} \frac{\delta b}{\chi} \simeq \frac{mN_{\rm DW}H_{\rm inf}}{2\pi\ell \langle\chi_{\rm inf}\rangle} \ .
\eeq
Since $\langle\chi_{\rm inf}\rangle$ is large as estimated in \EQ{Sinf}, the axion fluctuation is significantly suppressed, which was first pointed out by \cite{Linde:1991km}.

Assuming that the first oscillation induced by $V_{\cancel{\rm PQ}}$ does not contribute to the DM abundance, the dimensionless power spectrum is given by \cite{Kobayashi:2013nva}
\beq
\Delta^2_{\rm iso} \equiv \frac{k^3}{2\pi}\mathcal{P}_{\rm iso} \simeq \lmk\frac{\Omega_{\rm mis}}{\Omega_{\rm DM}}\frac{\del \Omega_{\rm mis}}{\del \ln \theta_{\rm amp}}\frac{\delta a}{f_a}\rmk^2,
\eeq
where one can estimate the amplitude of oscillation as $\theta_{\rm amp}\sim |\theta^{(\cancel{\rm PQ})}_{\rm min}-\theta_{\rm min}^{(\rm QCD)}|$.
The recent result of the Planck collaboration \cite{Planck:2018jri} puts an upper bound on the scale-invariant and uncorrelated isocurvature perturbation,
\beq
&&\beta_{\rm iso}(k_0) \equiv \frac{\mathcal{P}_{\rm iso}}{\mathcal{P}_{\zeta}} < 0.038 \ ,\\[1ex]
&\leftrightarrow& \Delta_{\rm iso}^2 < 8.3\times10^{-11} \ ,
\eeq
where $\mathcal{P}_{\zeta}$ represents the power spectrum of the adiabatic perturbations.

Neglecting the anharmonic effect, the upper bound on the inflation scale is estimated as
\beq
H_{\rm inf} &\lesssim& 2.5\times10^{15}\GeV\lmk\sqrt{\frac{c_S}{n}\frac{n!}{\lambda_S}}\rmk^{\frac{1}{n-2}}\nonumber\\
&\times& \lmk\frac{\ell}{mN_{\rm DW}}\rmk^{\frac{n-1}{n-2}} \lmk\frac{10^{-2}}{\Omega_{\rm mis}/\Omega_{\rm DM}}\rmk^{\frac{n-1}{n-2}}.
\eeq
For example, $H_{\rm inf}\lesssim 1.0\times10^{15}\GeV$ for $(N_{\rm DW},\ell,m,n) = (2,3,9,6)$, $\lambda_S=1$, $c_S=1$, and $\Omega_{\rm mis}/\Omega_{\rm DM}=10^{-2}$.
Considering the current bound on $H_{\rm inf}$, we are able to take $10^{-4} \lesssim \lambda_S \lesssim 1$.
Note that when $\delta'$ is finely tuned to the value where the potentials align with each other, the isocurvature bound is significantly enhanced by the anharmonic effect, as can be seen from \FIG{fig:abund}.

\subsection{Total abundance}

Including the contribution from the misalignment production, let us clarify the viable parameter space for solving the domain wall problem.
The total abundance of the axion DM is given by the sum of \EQ{Omegadec} and \EQ{small_l} or \EQ{large_l}, i.e.,
\beq
\Omega_{a,{\rm tot}} \simeq \Omega_{a,{\rm dec}} + \Omega_{a, {\rm mis}} \ .
\eeq
\FIG{fig:viable} describes the total axion abundance as a function of $|\lambda|$ with the red dotted contours of $f_a/(10^{10}\GeV) = 2^{-1/2}-2^{2}~(2^{3/2}-2^{7/2})$
for $\kappa=0.1 \, (0.5)$.
While the blue shaded region is excluded by the constraint of the backreaction (\ref{backreaction}), the string-wall system cannot collapse because $T_{\rm osc}\lesssim T_{\rm osc}^{(\rm conv)}$ in the orange shaded region.
We set $(N_{\rm DW},\ell,m,n)=(2,3,9,6)$, $\delta'=1$, $\lambda_S=10^{-4}$, and $m_S=10^{-20}\GeV$.
The total abundance is strongly dependent on the annihilation temperature $\kappa$.
For a given $f_a$, the maximum and minimum of $|\lambda|$ are determined from the backreaction and the stability of the string-wall system.
Although we need precise simulations, $f_a\lesssim 8\times10^{10}\GeV$ is required to avoid the overproduction of the axion for $\kappa\gtrsim0.5$.
However, if the string-wall system is very long-lived and $\kappa\ll\mathcal{O}(0.1)$, the upper bound on $f_a$ would be severer.

\begin{figure}[t!]
\centering
\includegraphics[width=8cm]{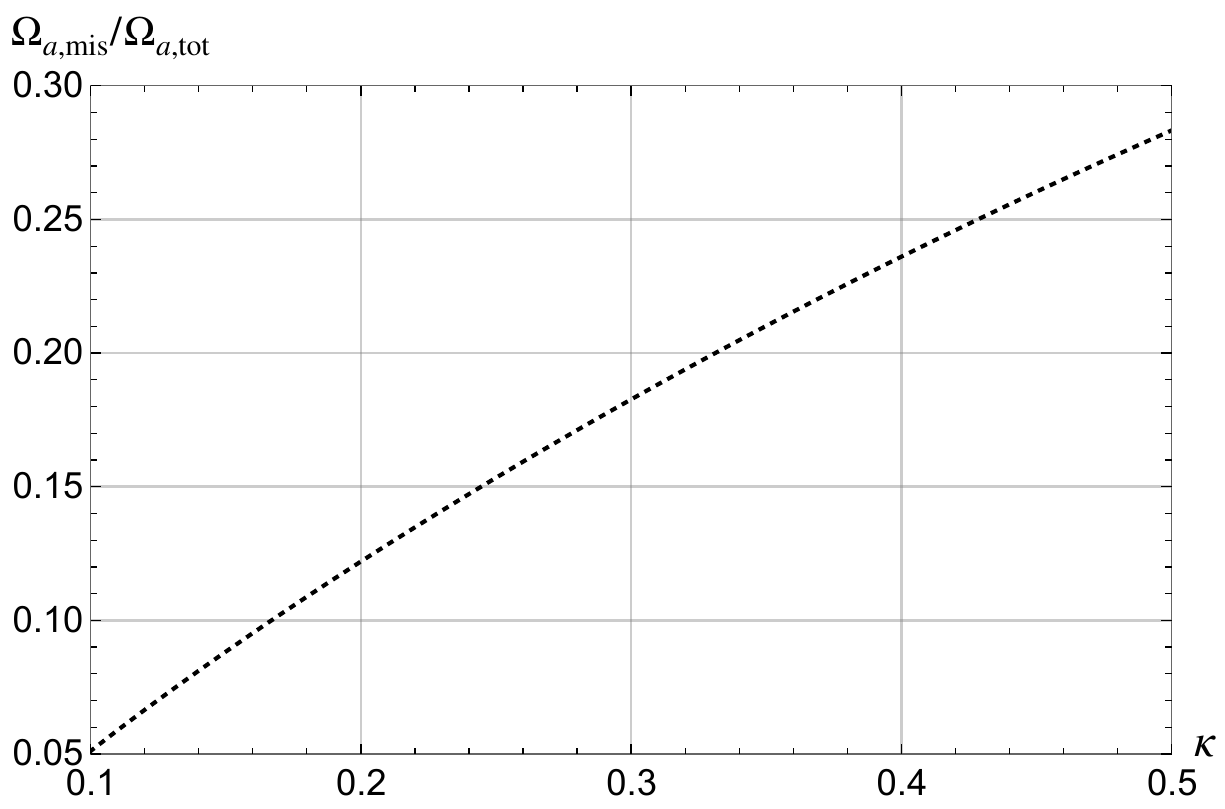}
\caption{
The fraction of the misalignment contribution $\Omega_{a,{\rm mis}}/\Omega_{a,{\rm tot}}$ as a function of the annihilation temperature $(\kappa)$.
We set $\delta'=1$, $\lambda_S=10^{-4}$,  $m_S=10^{-20}\GeV$, $(N_{\rm DW},\ell,m,n)=(2,3,9,6)$, $f_a=10^{10}\GeV$, and $\lambda=\lambda_{\rm max}$, which corresponds to the region of $\lambda=\lambda_{\rm max}$ in \FIG{fig:abund} for various values of $\kappa$.
}
\label{fig:ratio}
\end{figure}

Let us mention the relative importance of the contribution from the misalignment production.
In the conventional QCD axion scenario with a constant bias potential \cite{Kawasaki:2014sqa}, the fraction of the misalignment contribution is $\mathcal{O}(1)\%$ for $N_{\rm DW}=6$ and $f_a=10^{10}\GeV$, depending on the bias size.
In the presence of the extra potential, both contributions from the misalignment and the domain wall collapse can change.
However, while the effect of the trapped misalignment is very weak in our model as shown in \FIG{fig:abund},
it is still unclear when the string-wall network collapses (parameterized by $\kappa$), which strongly determines the contribution from the domain wall decay.
Therefore, the relative importance of the misalignment contribution dominantly depends on $\kappa$ in our current analysis.
In \FIG{fig:ratio}, we have shown the proportion of the misalignment contribution, $\Omega_{a, {\rm mis}}/\Omega_{a,{\rm tot}}$ as a function of $\kappa$ for $\lambda=\lambda_{\rm max}$ and $f_a=10^{10}\GeV$.
If the annihilation temperature $(\kappa)$ is relatively large, the misalignment axion production gives a relevant contribution, $\Omega_{a,{\rm mis}}/\Omega_{a,{\rm tot}} = \mathcal{O}(10)\%$.

\section{UV models
\label{sec:model}}

The scalar potential in Eq.~\eqref{VPQ} does not contain higher dimensional terms such as the terms with any lower powers of $|S|$ and the  $S$-$P$ mixing terms with other powers. To explain the specific potential form without fine tuning, we consider a supersymmetric version of the model.
The SUSY model for $\ell=1$ has been discussed in ref.~\cite{Ibe:2019yew}. Our focus here is to explore models for $\ell \geq 2$. In particular, we consider the case of $(\ell,m,n)=(2,9,6)$ as our benchmark.

Let us consider a superpotential that respects both the $U(1)_{\rm R}$ and $U(1)_{\rm PQ}$ symmetries,
\begin{align}
\label{eq:superpotential}
    W = X(P\bar{P} - v_{\rm PQ}^2)
    + \frac{\lambda_S}{6!} \frac{Y S^6}{\Mpl^4}
    + \lambda Z\left( \frac{P^2}{2!} + \frac{S^9}{9! \Mpl^7} \right) \ .
\end{align}
Here, \( X \), \( P \), \( \bar{P} \), \( S \), \( Z \), and \( Y \) are chiral superfields whose charges under the $U(1)_{\rm R}$ and $U(1)_{\rm PQ}$ are listed in Table~\ref{tab:charges}. The $F$-term components of \( X \), \( Y \), and \( Z \) generate the scalar potential given in Eq.~\eqref{VPQ}. The mass term for the scalar component of \( S \) arises from SUSY breaking effects. In addition to the terms shown above, higher dimensional terms in the superpotential such as  
\begin{align}
    W \supset X (P \bar{P})^2,\quad
    X S^9 \bar{P}^2,\quad
    Y S^6 P\bar{P},\quad
    Z P^3 \bar{P} ,
\end{align}
may also appear. However, these terms do not generate lower powers of \( |S| \), and they are sub-leading compared to the relevant $S$-$P$ mixing term.
Moreover, higher dimensional operators in the Kähler potential, such as  
\begin{align}
    K \sim \frac{1}{\Mpl^2} (|X|^2 |S|^2 + |Y|^2 |S|^2 + |S|^4) \ ,
\end{align}
do not induce lower dimensional terms for \( |S|^n \), as discussed in ref.~\cite{Ibe:2019yew}.
The smallness of the couplings \( \lambda_S \) and \( \lambda \) is  ’t Hooft natural because the phase rotation symmetries of $Y$ and $Z$ are enhanced in the limit of  \( \lambda_S \to 0\) and \( \lambda \to 0 \).
We can also explain the smallness by introducing some spontaneously broken symmetries as in the case of ref.~\cite{Froggatt:1978nt}

\begin{table}[!t]
    \centering
    \begin{tabular}{c|cccccc}
        & $X$ & $Y$ & $Z$ & $S$ & $P$ & $\bar{P}$ \\
        \hline
        $U(1)_{\rm R}$    & $2$ & $2$  & $2$  & $0$ & $0$  & $0$ \\
         $U(1)_{\rm PQ}$ & $0$ & $-12$ & $-18$ & $2$ & $9$ & $-9$ \\
    \end{tabular}
    \caption{Charge assignments of the chiral superfields under $U(1)_{\rm PQ}$ and $U(1)_{\rm R}$.}
    \label{tab:charges}
\end{table}

The small mass of \( S \), where typically we take it around \( m_S = 10^{-20}~\mathrm{GeV} \), requires significant fine tuning. In supersymmetric models, \( m_S \) is generally given by the order of the gravitino mass, \( m_{3/2} \sim \Lambda^2 / M_{\mathrm{Pl}} \), where \( \Lambda \) denotes the SUSY breaking scale. This mass scale is generated by the Kähler term of the form \( K \sim |T|^2 |S|^2 / M_{\mathrm{Pl}}^2 \) where \( T \) is the SUSY breaking field.
While the gravitino mass can be much smaller than the soft masses of sfermions and gauginos, achieving \( m_S = 10^{-20}~\mathrm{GeV} \) still demands severe fine tuning. 
This problem can be addressed if the SUSY breaking sector is sequestered from the $S$-$P$ sector
by e.g. using no-scale Kähler potential~\cite{Goto:1991gq}.
In this case, the scalar and fermionic superpartners of the axion, the {\it saxion} and {\it axino}, can be very light,
and their impact on cosmology is a non-trivial issue, which is left for a future study.

Lastly, let us also mention models with other combinations of \((\ell,m,n)\), including the case \((\ell,m,n) = (3,9,6)\), which corresponds to the scenario illustrated in Fig.~\ref{fig:large_l}. Similar to Eq.~\eqref{eq:superpotential}, one can construct a superpotential that generates both the \(|S|^{2n}\) term and the mixing term between \(P\) and \(S\). Such a superpotential generally respects the \(U(1)_{\rm PQ}\) and \(U(1)_{\rm R}\) symmetries. However, relatively lower dimensional operators of the form,
\begin{equation}
V \sim \frac{|S|^{2k_s} |\bar{P}|^{2k_p}}{M_{\text{Pl}}^{2k_s + 2k_p - 4}} \ ,
\end{equation}
with positive integers \(k_s, k_p\), may also be allowed. These terms can be problematic for our scenario if \(S\) acquires a large VEV, as they can shift the minimum of the QCD axion potential. A detailed analysis of such generic setups will be performed elsewhere.

\begin{figure*}[t!]
\begin{minipage}[t]{16.cm}
\centering
\includegraphics[width=5.5cm]{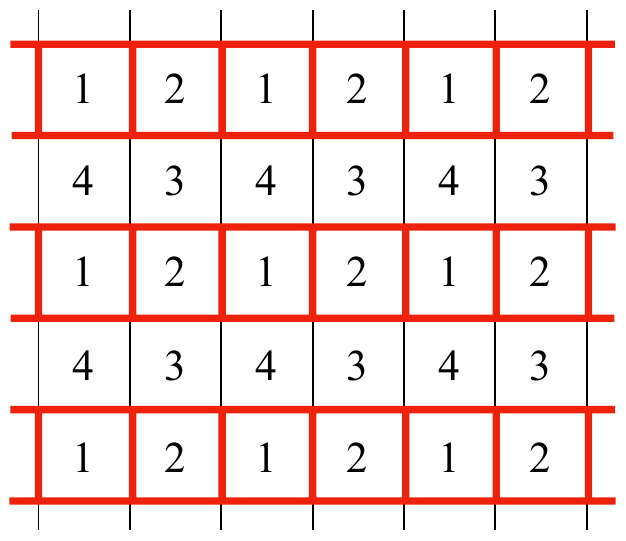}
\hspace{15mm}
\centering
\includegraphics[width=7.5cm]{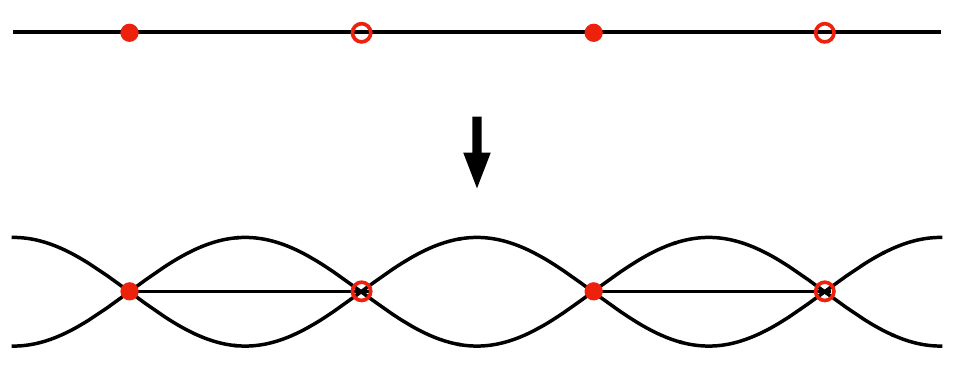}
\end{minipage}
\caption{
Left: the system for $\ell=4$, $N_{\rm DW} = 3$. 
The red solid lines represent the structure above the QCD scale.
Right: the system for $\ell=2, N_{\rm DW}=5$.
The red bullets and circles denote the strings and anti-strings, and the black line is the domain wall.
}
\label{fig:structure}
\end{figure*}

\section{Conclusions and Discussions
\label{sec:Discussion}}

We have explored the dynamics of a post-inflationary QCD axion coupled to a light scalar field, focusing on their interplay in resolving the domain wall problem and generating the axion DM. The spectator scalar field acquires a nonzero VEV during inflation, inducing an effective PQ-violating interaction that leads to an axion potential with multiple degenerate vacua below the PQ phase transition. This structure results in the formation of string-domain wall networks. At the QCD phase transition, a new potential contribution triggers the decay of domain walls before they dominate the Universe.
As the spectator scalar field relaxes to its true minimum, the effective PQ violation is turned off, realigning the axion potential with the QCD vacuum. We have followed the evolution of the axion-scalar system and studied axion DM production via domain wall decay and misalignment. Notably, we find that in some cases, the string-wall network can decay due to structural instability rather than volume pressure, enabling successful DM production with a higher axion decay constant than in conventional scenarios, without fine-tuning.

We have discussed how the string-wall system is destabilized by the structural instability for $(\ell,N_{\rm DW})=(3,2)$ in the presence of the extra potential $V_{\cancel{\rm PQ}}$.
Here we explore the other cases of $(\ell,N_{\rm DW})$.
For example, consider $(\ell,N_{\rm DW})=(4,3)$.
The initial network looks like lattice denoted by the black solid lines in the left panel of \FIG{fig:structure}.
At the QCD scale, one wall disappears and the ladder-like structure remains (red solid lines).
Although it looks unstable in the direction perpendicular to the ladder, one can see from the overall structure that it may be stable or very long-lived.
For $\ell < N_{\rm DW}$, e.g., $\ell=2,N_{\rm DW}=5$, the initial configuration is given by a sequential structure of one string attached by two domain walls, shown in the right panel of \FIG{fig:structure}.
Considering the scaling law, we expect that the original two domain walls are likely to be decomposed into two or three domain walls after $V_{\rm QCD}$ becomes dominant.
Since the part formed by three domain walls has stronger tension force, it shrinks so that the string and anti-string are pair-annihilated.
Still, it seems that a chain-like defect made up by two domain walls without strings remains stable. 
Although we conclude that the system is naively stable except for $(\ell,N_{\rm DW})=(3,2)$ and $f_a\lesssim 10^9\GeV$, the fate is still unclear and should be clarified by simulations.
Lastly, let us comment on another destabilization mechanism, where if the initial population of the axion field is biased, the structure can collapse immediately \cite{Gelmini:1988sf,Lalak:1993ay,Lalak:1993bp,Lalak:1994qt,Coulson:1995nv,Larsson:1996sp,Correia:2014kqa,Correia:2018tty,Krajewski:2021jje,Gonzalez:2022mcx,Kitajima:2023kzu}.
In most literature, $N_{\rm DW}=2$ case was studied because the percolation theory works.
If we identify the string-wall system created by the extra potential as the initial population, the stability of the system can be discussed.
Although there are several unclear points, e.g. how the case of $N_{\rm DW}\neq2$ is investigated in a percolation theory-like way, it may be an interesting direction for a future study.

\begin{figure}[t!]
\centering
\includegraphics[width=8cm]{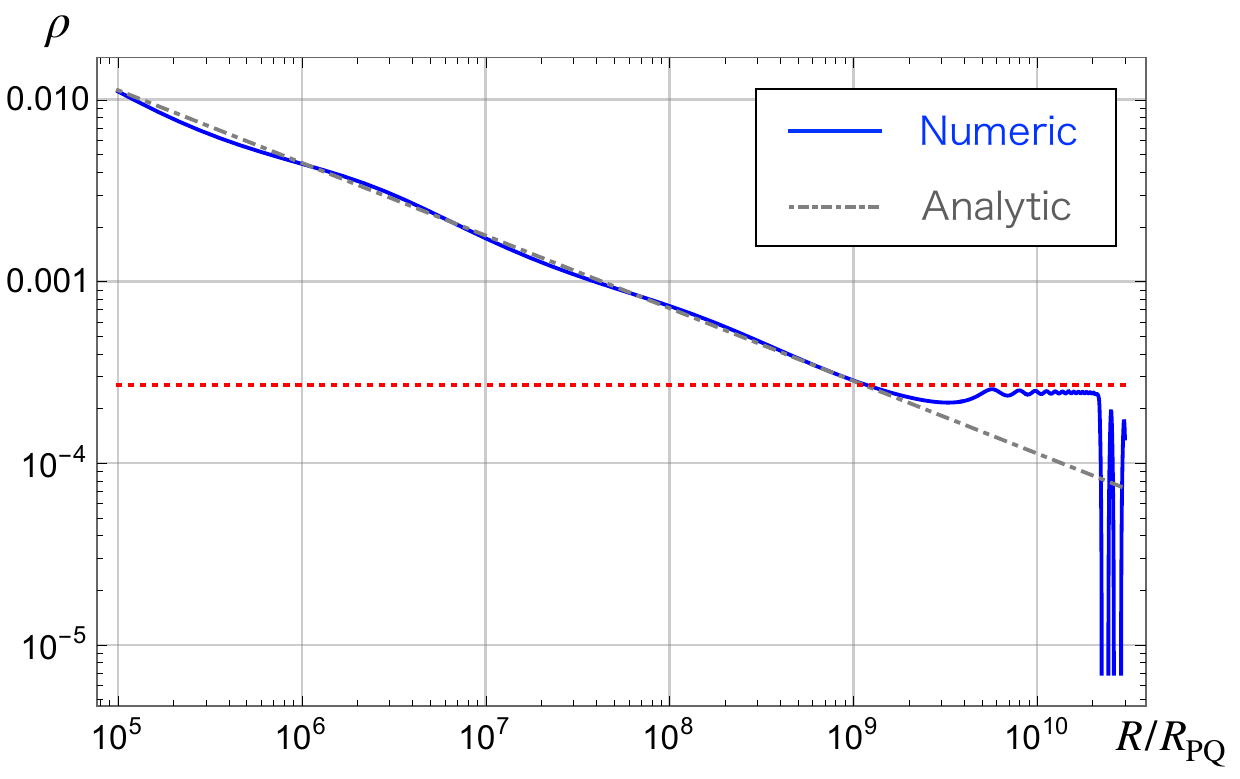}
\caption{
The time evolution of the real spectator field for $(N_{\rm DW},\ell,m,n)=(1,2,9,6)$.
We set $|\lambda|=10^{-8}$ and $\lambda_S=10^{-4}$, $m_S=10^{-20}\GeV$, $f_a=10^{10}\GeV$, and $\delta=-\pi/5$.
The red dotted line corresponds to \EQ{rhoconst}.
The oscillation begins at the QCD scale, $T\simeq v_{\rm PQ}\cdot(R_{\rm PQ}/R)\sim \Lambda_{\rm QCD}$.
}
\label{fig:real}
\end{figure}

In \SEC{sec:full}, we have followed the full equations of motion to reveal the backreaction and have found it possible to consider the region where the backreaction is non-negligible.
For a large $\lambda$, the trapped misalignment is more remarkable, and the abundance is significantly enhanced.
The estimation of the abundance was made in ref.~\cite{Jeong:2022kdr}, and we can expect that there is an upper bound on $|\lambda|$, because the abundance becomes independent of $f_a$ and increases with the strength of the trapping effect. 
In addition, since the phase $b$ of $S$ is affected by the extra potential above the QCD scale, it also gets involved with the $a$ evolution, which can alter the final abundance via the misalignment mechanism, depending on the parameters.
Although the detailed consequences will be studied in subsequent papers, let us mention one more interesting implication for the system.
The reason why $S$ never oscillates is that the sign of the mixing term is necessarily negative.
One potential way to flip the sign is to consider a real spectator field or to replace the mixing term $S^mP^\ell$ with $|S|^mP^\ell$.
In this case, the cosine term in the mixing term changes from negative to positive when the axion starts to oscillate by $V_{\rm QCD}$, i.e. $\cos(\ell\theta_{\rm min}^{(\cancel{\rm PQ})}+\delta)\rightarrow \cos\delta$ with $|\delta|<\pi/2$, as long as $N_{\rm DW}$ is equal to $1$ or a number which is not prime with $\ell$.
In \FIG{fig:real}, we show the time evolution of $S$ for $(N_{\rm DW},\ell,m,n)=(1,2,9,6)$ and $|\lambda|=10^{-8}$ which is over the bound (\ref{backreaction}), and one can find that $S$ oscillates at the QCD scale.
In that case, the oscillation amplitude of $S$ is larger for larger $|\lambda|$, and the oscillation energy would be dominant.
If we introduce a coupling of $S$ with light fermions to make $S$ decay into them, the energy remains as dark radiation which can be probed by future experiments, e.g. \cite{Sehgal:2019ewc,CMB-HD:2022bsz}, leading to another upper bound on $|\lambda|$.

Let us discuss the effect of the mixing term on the neutron EDM.
When the Hubble parameter becomes comparable to $m_S$, the spectator field starts to oscillate around the origin.
The current oscillation amplitude is estimated by 
\beq
\frac{|S(T_0)|}{f_a} \simeq \sqrt{\frac{s_0}{s_{S_{\rm osc}}}}\frac{|S(T_{S_{\rm osc}})|}{f_a} \simeq 10^{-16} \lmk\frac{10^{10}\GeV}{f_a}\rmk \ , 
\eeq
where $T_{S_{\rm osc}}$ denotes the temperature that $S$ starts to oscillate, $s_0, s_{S_{\rm {osc}}}$ are the entropy densities estimated at $T=T_0, T_{S_{\rm osc}}$, respectively, and we set $m_S=10^{-20}\GeV$ and $(\ell,m,n)=(2,9,6)$.
This induces the deviation from the CP conserving minimum, $\delta\theta_{\rm min} \simeq 10^{-175}$, which is negligibly tiny compared to the current bound on the static EDM.
In this class of models, the axion minimum at present can oscillate due to the much longer timescale of $S$'s oscillation compared to that of the axion.
While the amplitude of $S$ may be too small for the effect to be observable, the time-dependent EDM nevertheless presents an intriguing possibility in principle \cite{Graham:2011qk}.

\section*{Acknowledgments}

We would like to thank Qianshu Lu, Yu-Cheng Qiu, Kiwoon Choi, and Fuminobu Takahashi for fruitful discussions.
Y.N. is supported by Natural Science Foundation of Shanghai.
M.S. is supported by the MUR projects 2017L5W2PT.
M.S. also acknowledges the European Union - NextGenerationEU, in the framework of the PRIN Project “Charting unexplored avenues in Dark Matter” (20224JR28W).

\appendix

\section{Detailed analysis of axion evolution
\label{app:adiabatic}}

In this Appendix, we discuss the detail of the axion evolution.
In particular, our assumptions in the calculation of the misalignment production are confirmed.
Here we fix the parameters $(N_{\rm DW},\ell, m ,n) = (3,2,9,6)$, $\lambda_S=10^{-4}$, and $f_a=10^{10}\GeV$ as an example.

\subsection{Approximation by kinetic energy}

\begin{figure}[t!]
\centering
\includegraphics[width=8cm]{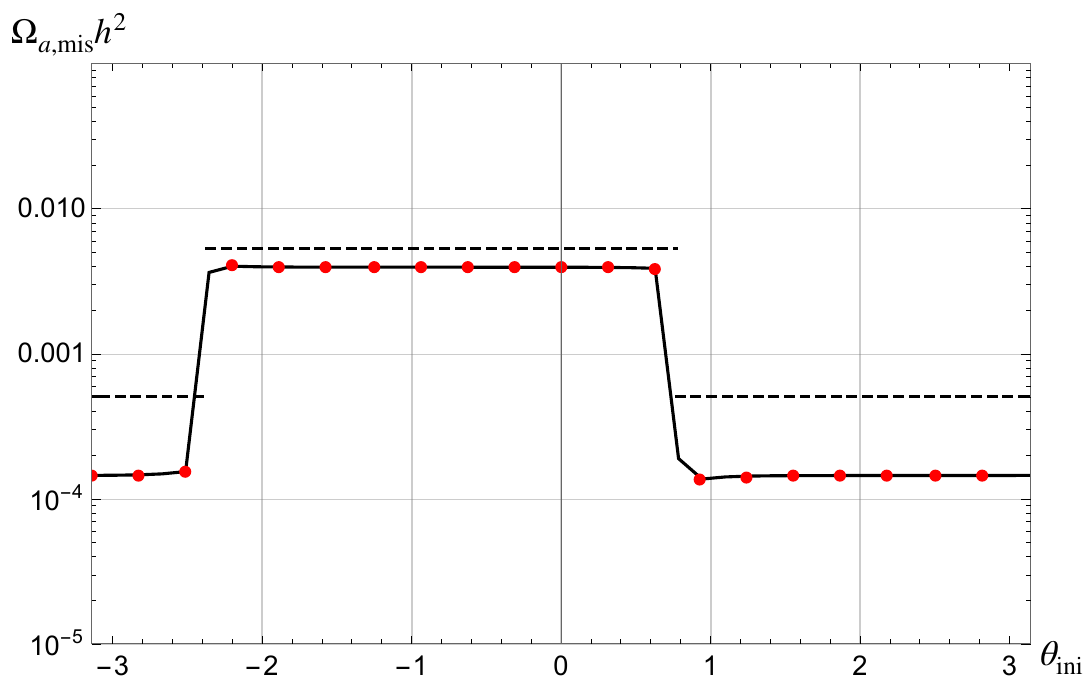}
\caption{
The dependence of the abundance produced by the misalignment mechanism on $\theta_{\rm ini}$.
We take $\lambda=\lambda_{\rm max} \simeq 4\times10^{-12}$ and $\delta'=1.6$.
The black line is the numerical results under our assumption, while the red points denotes the numerical results with much more precision.
The dashed line is the analytic result (\ref{analytic}).
}
\label{fig:thini}
\end{figure}

To save time for the numeric calculations, we adopt an approximation that the abundance can be estimated by the average of the kinetic energy after the axion oscillates around the current vacuum.
Since the extra potential $V_{\cancel{\rm PQ}}$ contributes to the negative potential energy, the number density increases or even the average of the kinetic energy slightly varies with time.
However, the change in the kinetic energy is negligibly small, because the time scale of the variation of vacuum energy is much slower than the oscillation scale.
In \FIG{fig:thini}, we show the abundance as a function of $\theta_{\rm ini}$ for $|\lambda| = \lambda_{\rm max}=4\times10^{-12}$ and $\delta'=1.6$.
The black line denotes the numerical results under our assumption, while the red points are given by much more precise numerical calculation.
The latter is estimated by the number density, not by the kinetic energy.
We have checked the number density is converged at the $1\%$ level.
One can see from this results that the estimation by the average of the kinetic energy is consistent with that by the number density at least $1\%$ level.

\subsection{Dependence on $\theta_{\rm ini}$}

\begin{figure}[t!]
\centering
\includegraphics[width=8cm]{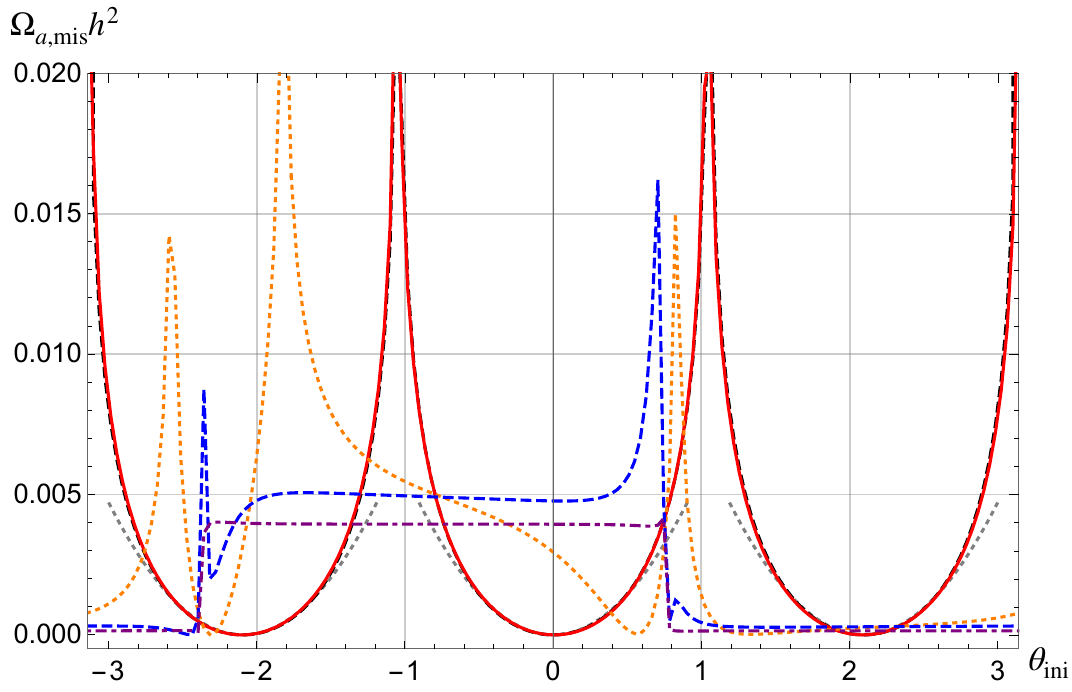}
\caption{
The dependence of the abundance $\Omega_{a,{\rm mis}}h^2$ on $\theta_{\rm ini}$ for $\delta'=1.6$ and various values of $|\lambda|$.
We show the numerical results for $|\lambda|=0$ (black dotted), $10^{-15}$ (red solid), $10^{-13}$ (orange dotted), $4\times10^{-13}$ (blue dashed), and $4\times10^{-12}$ (purple dot-dashed).
The gray dotted line denotes the analytic result in the vicinity of the minimum of $V_{\rm QCD}$.
}
\label{fig:thini2}
\end{figure}

Next let us discuss the important assumption of the dependence of the abundance via the misalignment mechanism on $\theta_{\rm ini}$.
\FIG{fig:thini} also shows that the abundance $\Omega_{a,{\rm mis}}h^2$ does not depend on $\theta_{\rm ini}$.
Note that $\theta=(\pm\pi-\delta')/\ell\simeq -2.37,0.77$ correspond to the potential top of $V_{\cancel{\rm PQ}}$.
Around there, the onset of the oscillation is significantly delayed due to the anharmonic effect, and the abundance slightly changes for different $\theta_{\rm ini}$.
Except for such regions, this numerical results clarify that the misalignment production is determined by which minimum the axion initially starts to oscillate around in the case of $|\lambda|=\lambda_{\rm max}$.

In \FIG{fig:thini2}, we summarize the numerical results of $\Omega_{a,{\rm mis}}h^2$ for various $|\lambda|$.
The black dotted, red solid, orange dotted, blue dashed, and purple dot-dashed lines are respectively for $|\lambda| = 0, 10^{-15}, 10^{-13}, 4\times10^{-13}$, and $4\times10^{-12}$.
While the extra potential does not affect the axion evolution for $|\lambda|=10^{-15}$, as $|\lambda|$ increases, one can see, from the result for $|\lambda|=10^{-13}, 4\times10^{-13}$ which gives $T_{\rm osc}$ comparable to $T_{\rm osc}^{(\rm conv)}$, that the anharmonic peaks move and will be finally converged to the potential top of $V_{\cancel{\rm PQ}}$. 
This implies that our assumption of the independence of $\theta_{\rm ini}$ is guaranteed,
when $|\lambda|$ is conservatively $10$ times larger than the threshold value ($\lambda_{\rm th}\approx  10^{-13}$) at $T_{\rm osc}=T_{\rm osc}^{(\rm conv)}$:
\beq
\mathcal{O}(10)\lambda_{\rm th} \lesssim |\lambda| \lesssim \lambda_{\rm max}. 
\label{indep_cond}
\eeq
Below, we further discuss the $\theta_{\rm ini}$ independence.

\subsection{Violation of adiabaticity}

Finally, we would like to clarify why the dependence on $\theta_{\rm ini}$ disappears in the range of (\ref{indep_cond}).
Our scenario is a kind of the trapped misalignment, which can be classified into the two regimes (pseudo-smooth shift or trapped). 
In the conventional smooth-shift scenario, it is assumed that the trapping effect is sufficiently strong,
so that the final axion abundance is directly linked to the initial misalignment angle.
In our model, however, the trapping effect decreases with time and becomes very weak around the transition,
$T \sim T_{\rm tr}$.
As a result, the axion evolution is not fully adiabatic, which motivates us to estimate the axion abundance
by averaging over all possible minima of $V_{\cancel{\rm PQ}}$.

\begin{figure*}[t!]
\begin{minipage}[t]{16.cm}
\centering
\includegraphics[width=7.8cm]{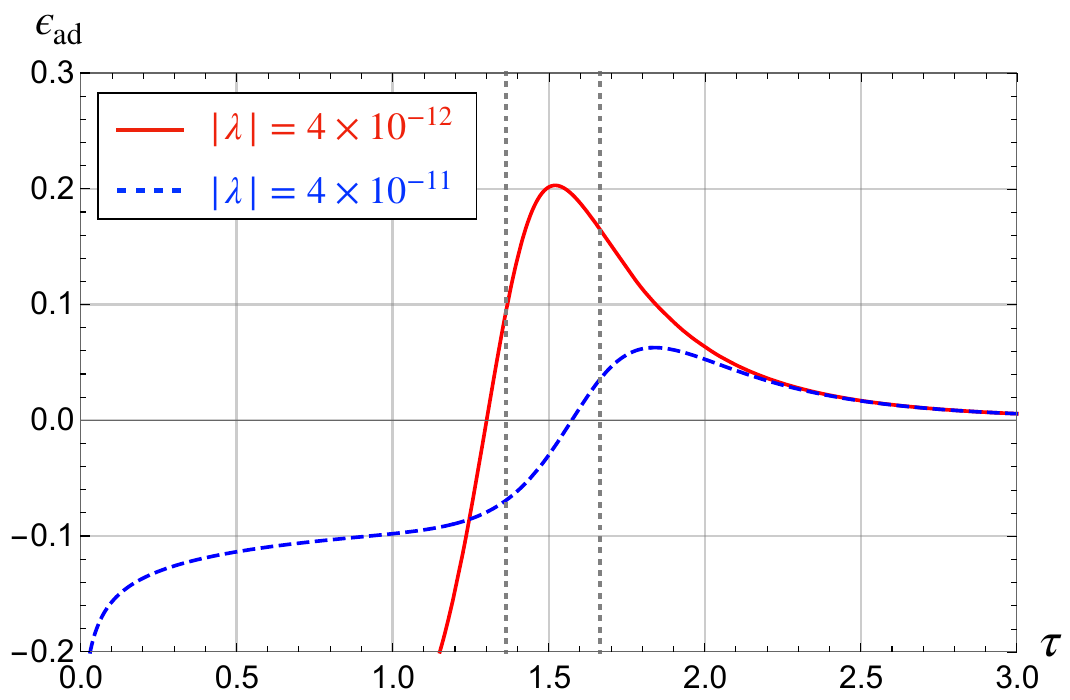}
\hspace{2mm}
\centering
\includegraphics[width=7.8cm]{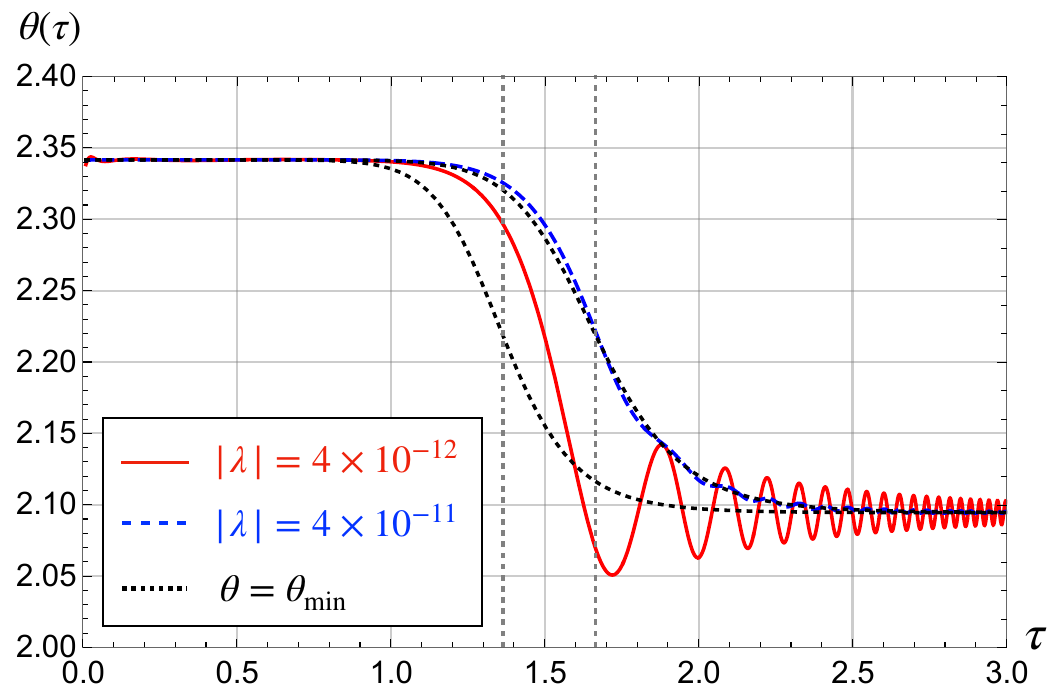}
\end{minipage}
\caption{
Left: the evolutions of $\epsilon_{\rm ad}$ as a function of $\tau\equiv T_{\rm osc}^{(\rm conv)}/T$ for $|\lambda|=4\times10^{-12}$ $(4\times10^{-11})$, denoted by the red solid line (the blue dashed line).
Here we take $\delta'=1.6$ and $\theta_{\rm ini}=2$.
The gray dotted vertical lines represent the timing of $m_{\rm QCD}=m_{\cancel{\rm PQ}}$ for $|\lambda|=4\times10^{-12},4\times10^{-11}$ from the left.
Right: the corresponding field evolution of the axion. 
The black dotted lines are the potential minimum. 
}
\label{fig:adiabaticity}
\end{figure*}

\begin{figure}[t!]
\centering
\includegraphics[width=8cm]{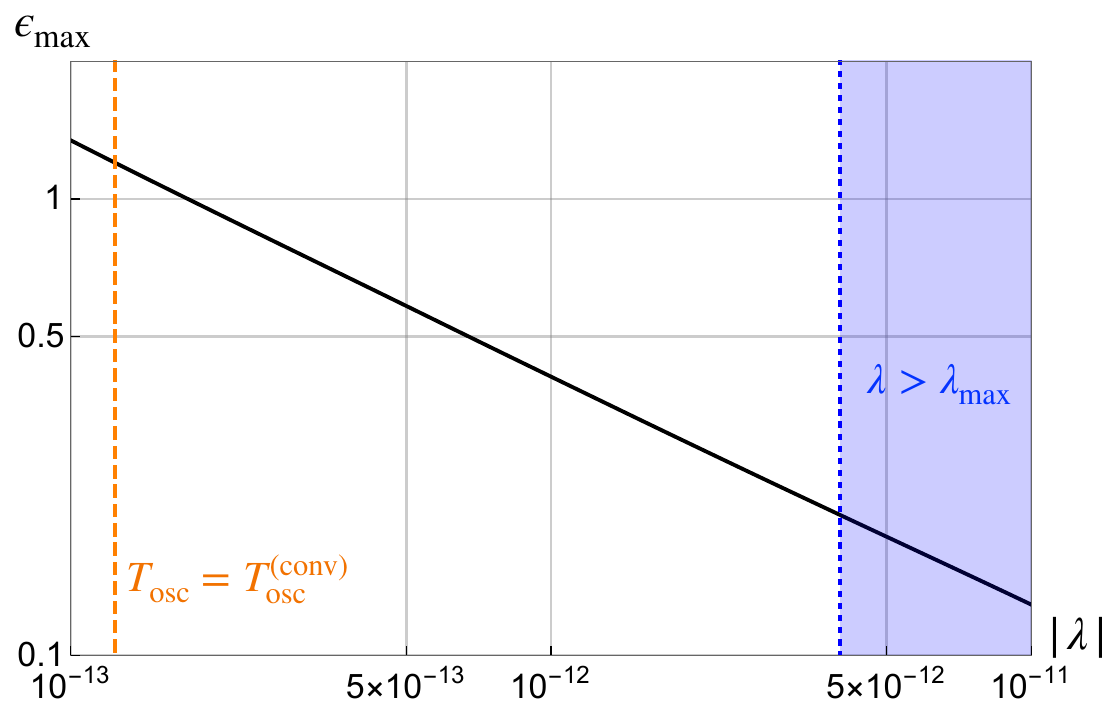}
\caption{
The maximum value of $\epsilon_{\rm ad}$ as a function of $|\lambda|$ for $\delta'=1.6$.
Here we consider the case that the axion initially starts to oscillate at the minimum, $\theta = (2\pi-\delta)/\ell$.
The violation of the adiabaticity is evaluated at the temporal position of the axion field.
The blue shade region is excluded by the backreaction (\ref{backreaction}).
}
\label{fig:max_e}
\end{figure}

To understand the physics more clearly, let us introduce a function indicating violation of adiabaticity,
\beq
\epsilon_{\rm ad}(\theta_*, T) \equiv \frac{\dot{m}_{\rm eff}}{m_{\rm eff}^2},
\eeq
where $\theta_*$ is taken to be the temporal field value of the axion, and we define the effective mass as
\begin{align}
m_{\rm eff} \equiv \sqrt{\left|\frac{\del^2 (V_{\rm QCD}+V_{\cancel{\rm PQ}})}{\del a^2}\right|}.
\end{align}
Substituting the potentials (\ref{VQCD}) and (\ref{PQVpotential}), we obtain the violation of adiabaticity,
\begin{align}
\epsilon_{\rm ad}(\theta_*,T) = &\frac{H}{m_{\rm eff}^3}[\tilde{b}m_a^2\cos(N_{\rm DW}\theta_*)\nonumber\\
&-\frac{m}{n-1}m_{\cancel{\rm PQ}}^2 \cos(\ell \theta_*+\delta)].
\end{align}
The function quantifies the degree of the time evolution of the potential shape around the temporal axion field ($\theta=\theta_*$).
In other words, for smaller $\epsilon_{\rm ad}$, the potential changes more slowly, so that the axion can track the potential minimum.

In \FIG{fig:adiabaticity}, we show the violation of the adiabaticity and the corresponding axion evolution as a function of time $\tau\equiv T_{\rm osc}^{(\rm conv)}/T$ for $\delta'=1.6$.
The red solid (blue dashed) line is the result for $|\lambda|=4\times10^{-12}~(4\times10^{-11})$.
When $m_a\simeq m_{\cancel{\rm PQ}}$ (denoted by the gray vertical lines), the adiabaticity is most strongly violated, and we define it as $\epsilon_{\rm max}$.
We find that the adiabaticity is violated enough even for the maximum of $|\lambda|$,\footnote{We do not know how small $\epsilon_{\rm ad}$ should be sufficient for strong adiabaticity, but typically, we can take $\epsilon_{\rm max} < 0.1$.} and the secondary oscillation is induced. 
Although we cannot take a larger value, if it was allowed, the axion evolution is so adiabatic that any extra oscillation is not induced (see the blue dashed line in \FIG{fig:adiabaticity}).

\FIG{fig:max_e} shows the maximum of $\epsilon_{\rm ad}$ as a function of $|\lambda|$ for $\delta'=1.6$.
One can see from this figure that the axion evolution is not adiabatic in the whole parameter range we are interested in.
Near the bound $T_{\rm osc}=T_{\rm osc}^{(\rm conv)}$ (the orange dashed line), the vacuum is being deformed when the axion starts to oscillate, and the abundance depends nontrivially on $\theta_{\rm ini}$.
On the other hand, in the range far from the orange dashed line, the secondary oscillation amplitude is induced dominantly (see also the right panel of \FIG{fig:adiabaticity}), which determines the final abundance of the axion.

\begin{figure}[t!]
\centering
\includegraphics[width=8cm]{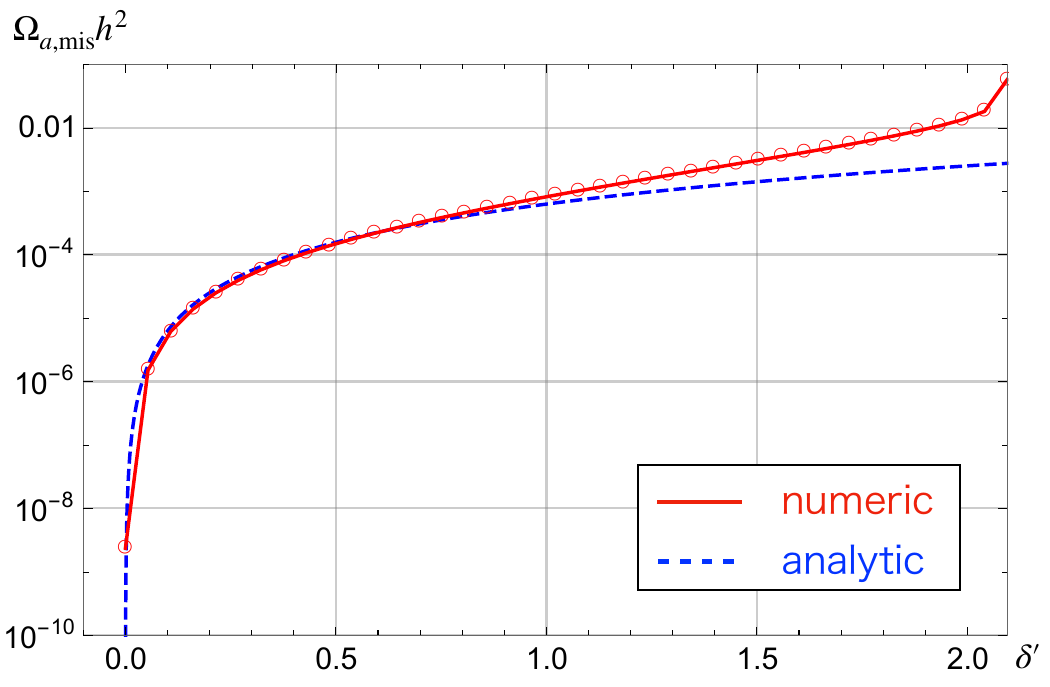}
\caption{
The dependence of $\Omega_{a,{\rm mis}}h^2$ on $\theta_{\rm ini}$ for $\lambda=\lambda_{\rm max}$.
The initial position of the axion field is taken so that the first oscillation amplitude is the unity, $\theta_{\rm ini} = -\delta'/\ell+1$.
The red circles are numerical results, while the blue dashed line corresponds to the analytical results, where we adjust the onset of oscillation with the numerical results.  
}
\label{fig:delta}
\end{figure}

Assuming that the secondary oscillation is dominant, we can estimate the abundance analytically. 
Since the oscillation amplitude is fixed by the phase difference between the minima of $V_{\cancel{\rm PQ}}$ and $V_{\rm QCD}$, the abundance is given by
\beq
\Omega_{\rm analy}h^2 = m_{a0}\frac{m_a(T'_{\rm osc}) a_{\rm amp}^2/2}{s(T'_{\rm osc})} \frac{s_0}{\rho_ch^{-2}}.
\label{analytic}
\eeq
Here $T'_{\rm osc}$ is the temperature at the onset of the oscillation, and can be typically $T_{\rm osc}$ but slightly different.
This formula reproduces no dependence on $\theta_{\rm ini}$.

In \FIG{fig:thini}, we show the analytic result by the black dashed line.
Here we take $T'_{\rm osc}=T_{\rm osc}$, and it is consistent with the numeric result up to an order one factor.%
\footnote{
The difference between the numerical result and \eqref{analytic} becomes more pronounced in the outer region, including large $\theta_{\rm ini}$. In this regime, the minimum of $V_{\cancel{PQ}}$ lies near the maximum of $V_{\rm QCD}$, leading to anharmonic effects. These effects modify the onset of oscillation, such that $T’_{\rm osc}$ becomes earlier than $T_{\rm osc}$, resulting in a suppressed abundance in the black solid line and red points. Nevertheless, \eqref{analytic} still provides a reliable order-of-magnitude estimate.
}
\FIG{fig:delta} shows the dependence of $\Omega_{a,{\rm min}}h^2$ on $\delta'$ for $|\lambda|=4\times10^{-12}$.
The red line is the numeric result, and the blue dashed line is the analytic result.
Here $T'_{\rm osc}$ is matched with the numerical estimation.
In the range of $\delta'\lesssim1$, they are consistent, but the anharmonic effect makes the difference for the outer range.

We have found the consistency between numeric and analytic results which is understandable from the viewpoint of adiabaticity.\footnote{Although we would see different conditions which guarantee our assumptions for different parameter sets of $(\ell,m,n)$, the physical behavior in the system is the same.}
Therefore, we have concluded that the axion abundance is dominated by the secondary oscillation, independent of $\theta_{\rm ini}$, because of the violation of adiabaticity.

\bibliography{reference}

\end{document}